\documentclass[usenatbib]{mn2e}
\usepackage{epsfig} 
\usepackage{graphicx}

\usepackage{url} 
\usepackage{amsmath}
\usepackage{graphicx}
\usepackage{ulem}

\def\dout{\bgroup
 \markoverwith{\lower-0.2ex\hbox
 {\kern-.03em\vbox{\hrule width.2em\kern0.45ex\hrule}\kern-.03em}}%
 \ULon}
\MakeRobust\dout

\newcommand{\K}{\mathrm{K}}

\newcommand{\cm}{\;\mathrm{cm}}

\newcommand{\s}{\;\mathrm{s}}
\newcommand{\erg}{\;\mathrm{erg}}
\newcommand{\km}{\;\mathrm{km}}

\newcommand{\Myr}{\;\mathrm{Myr}}
\newcommand{\Gyr}{\;\mathrm{Gyr}}

\newcommand{\Hz}{\;\mathrm{Hz}}

\newcommand{\Mpc}{\;\mathrm{Mpc}}

\newcommand{\Msol}{\; \mathrm{M}_{\odot}}

\newcommand{\apj}{ApJ}
\newcommand{\apjl}{ApJ}
\newcommand{\apjs}{ApJS}
\newcommand{\aap}{A$\&$A}
\newcommand{\CQG}{Class. Quantum Gravity}
\newcommand{\araa}{ARAA}

\newcommand{\mnras}{MNRAS}

\newcommand{\aj}{AJ}
\newcommand{\nat}{Nature}
\newcommand{\nar}{New Astron. Rev.}
\newcommand{\physrep}{Physics Reports}
\def\gta{\mathrel{\spose{\lower 3pt\hbox{$\mathchar"218$}}
        \raise 2.0pt\hbox{$\mathchar"13E$}}}

\defcitealias{TH09}{TH09}

    \def\beq{\begin{equation} }
    \def\eeq{\end{equation} }
    \def\spose#1{\hbox to 0pt{#1\hss}}
    \def\ltsim{\mathrel{\spose{\lower.5ex\hbox{$\mathchar"218$}}
     \raise.4ex\hbox{$\mathchar"13C$}}}

\title[IGM global warming by high-redshift miniquasars]
{X-ray emission from high-redshift miniquasars:\\
self-regulating the population of massive black holes through global warming}
\author[T.Tanaka, R.Perna and Z.Haiman]
{
Takamitsu Tanaka$^{1}$\thanks{E-mail: taka@mpa-garching.mpg.de}, Rosalba Perna$^{2}$, and Zolt\'an Haiman$^{3}$
\\$^{1}$Max-Planck-Institut f\"ur Astrophysik, Karl-Schwarzschild-Str. 1, 85741 Garching, 
Germany
\\$^{2}$JILA and Department of Astrophysical and Planetary Science,
University of Colorado at Boulder, 440 UCB, Boulder, CO, 80309, USA
\\$^{3}$Department of Astronomy, Columbia University, 550 West 120th Street, New York, NY 10027, USA
}

\begin{document}

\maketitle

\label{firstpage}
\begin{abstract}
Observations of high-redshift quasars at $z\gta 6$ imply that
supermassive black holes (SMBHs) with masses $M\gta10^{9}\Msol$ were
in place less than $1 \Gyr$ after the Big Bang. If these SMBHs
assembled from ``seed'' BHs left behind by the first stars, then they
must have accreted gas at close to the Eddington limit during a large
fraction ($\gta 50\%$) of the time.  A generic problem with this
scenario, however, is that the mass density in $M\sim 10^{6}\Msol$
SMBHs at $z\sim 6$ already exceeds the locally observed SMBH mass
density by several orders of magnitude; in order to avoid this
overproduction, BH seed formation and growth must become significantly
less efficient in less massive protogalaxies through some form of feedback,
while proceeding unabated in the most massive galaxies that formed first.  Using
Monte-Carlo realizations of the merger and growth history of BHs, we
show that X-rays from the earliest accreting BHs can provide such a
feedback mechanism, on a global scale.
Our calculations paint a self-consistent picture
of {\it black-hole-made climate change}, in which the first
miniquasars---among them the ancestors of the $z\sim 6$ quasar
SMBHs---globally warm the intergalactic medium
and suppress the formation and growth of
subsequent generations of BHs.  We present two specific models with
global miniquasar feedback that provide excellent agreement with
recent estimates of the $z=6$ SMBH mass function.  For each of these
models, we estimate the rate of BH mergers at $z>6$ that could be
detected by the proposed gravitational-wave observatory {\it
eLISA}/NGO.
\end{abstract}

\section{Introduction}
\label{sec:intro}

The discovery of bright quasars at redshifts $z\gta 6$ in the Sloan
Digital Sky Survey \citep[see the review by][]{Fan06}, the
Canada-France High-z Quasar Survey \citep{Willott+10a}, and the
current redshift record-holder at $z=7.08$ in the UKIDSS
\citep{Mortlock+11} indicates that supermassive black holes (SMBHs) as
massive as several $10^9{\rm M_\odot}$ were already in place when the
Universe was less than $1$ Gyr old.

The mechanism by which these early massive BHs formed and grew remains
poorly understood (see, e.g., \citealt{Haiman12} for a recent
comprehensive review).  One class of explanations is the rapid
collapse of primordial gas into a $10^4-10^6~{\rm M_\odot}$ black
hole, either directly (e.g. \citealt{HR93,LoebRasio94}), by accreting
onto a pre--existing smaller seed BH (e.g. \citealt{VolRees05}), or by
going through the interim state of a very massive star
(e.g. \citealt{Bond+84}), a rapidly accreting massive ``proto-star''
\citep{Begelman+06,Hosokawa+12} or a dense stellar cluster
\citep{Omukai+08,DV09}.  These models rely on rapid gas contraction in
deep potential wells (halos with virial temperatures $T_{\rm vir}\gta
10^4$K;
\citealt{OH02,BrommLoeb03,LodatoNatarajan06,SpaansSilk06,WA08,RH09b,SBH10})
and require the gas to avoid fragmenting early on by cooling via ${\rm
H_2}$ and/or metals.  It is unclear whether this special
configuration---warm, dense, metal-free gas in relatively massive halos---is
realized in nature \citep{Dijkstra+08,SSG10,Petri+12,IO12, Agarwal+12}.

An alternative possibility, which we explore further in this paper, is
that metal--free stars, with masses $\sim 10-100\Msol$, form at
redshifts as high as $z\gta 25$ \citep{Abel+02,Bromm+02,Yoshida+08},
leave behind remnant BHs with similar masses \citep{Heger+03}, and
subsequently grow by mergers and by mass accretion near the Eddington
limit (\citealt{HaimanLoeb01}; \citealt{Haehnelt03};
\citealt{Bromley+04}; \citealt{YooMirald04}; \citealt{Taniguchi04};
\citealt{Shapiro05}; \citealt{VolRees06}; \citealt{TH09}, hereafter TH09).
The primary theoretical uncertainty of this scenario is
whether the seed BHs can sustain such high accretion rates.

To reach masses of $>10^9\Msol$ by $z\sim 7$, the stellar-mass BHs
must grow near the Eddington limit without significant interruption
(with a mean duty cycle $\gta 0.5$ at radiative efficiency $0.07$;
\citetalias{TH09}).\footnote{Alternatively, the accretion could occur
intermittently in brief episodes of super-Eddington accretion.}
However, the accretion rates of BHs are limited by their gaseous
environments: the rates can be significantly sub-Eddington if the seed
mass is light, or if the BH is embedded in a low--density and/or
high--temperature medium, as may often be the case at high redshifts
(see \S~\ref{subsubsec:accretion}).
Another limitation is radiative feedback: recent work has shown that
the accretion is episodic, with time-averaged Eddington duty cycle of
at most $\sim1/3$ (\citealt{Milos+09,PR12}; see also earlier work by
\citealt{CO01} for radiative feedback forcing accretion to be episodic
with a low duty cycle at lower redshifts).  This local feedback could
be mitigated if the flow is non-spherical (i.e., in a disk) and
the radiation escapes vertically without compromising the equatorial
fuel supply.
Finally, a third type of limitation could be the lack of continuous
fuel supply on larger scales.  Interestingly, \cite{DiMatteo+12} have
found that, at least for $\ga 10^{5}\Msol$ BHs residing in massive
($\ga 10^{9}\Msol$) galaxies at lower redshifts $6<z<12$, filamentary
accretion of cool gas into the host galaxy may deliver gas to the
central regions of massive galaxies near the rates corresponding to
the BH's Eddington limit.  The large Eddington ratios ($\sim 1$;
\citealt{Willott+10b}) and duty cycles ($\ga 0.5$;
\citealt{Shankar+10}) inferred from observations of the quasar SMBHs
at $z\sim 6$ appear to be consistent with a sustained and rapid BH
growth.

If the BHs can indeed maintain such high accretion rates, then they
can grow into $\gta 10^9 {\rm M_\odot}$ SMBHs by $z>6$.  {\it However,
another, less appreciated problem then arises: these optimistic
assumptions inevitably lead to a severe over-production for the space
density of lower-mass} ($10^{5-7} {\rm M_\odot}$) {\it BHs}.  The
global comoving mass density of SMBHs in galactic nuclei in such
models can exceed the locally observed mass density by several orders
of magnitude, even if less than $0.1\%$ of star-forming halos
form a seed BH \citep{Bromley+04, TH09}.
The overproduction can be avoided if the growth of lower-mass
nuclear BHs is suppressed at late times, e.g. by
imposing an early $M-\sigma$ relation or scaling the
BH growth with the host galaxy merger history.

The purpose of the present paper is to consider an alternative
solution to this problem.
In the SMBH growth models of \citetalias{TH09}, the
$z\sim 6$ SMBHs originate from seeds born in the first minihalos at
redshifts $z\gta 25$.  On the other hand, the lower-mass ($M\ltsim
10^{7} {\rm M_\odot}$) BHs arise primarily from seeds born inside
minihalos collapsing later, at $z\sim 15$.  In principle, therefore,
the overproduction of the $\gta 10^{5-7} {\rm M_\odot}$ BHs can be
avoided, provided that the seed BHs are either unable to form, or
unable to grow, in the vast majority of minihalos at $z\ltsim 20$.  A
natural reason for this could be photoionization heating of the intergalactic medium
(IGM) by the earliest accreting seed BHs themselves.  The X-rays emitted from
these holes have a long mean free path and establish an early X-ray
background, which can pre-heat and pre-ionize the IGM \citep{Oh01,
Venkat+01, RicOst04, Madau+04}.  Once this heating sufficiently
elevates the IGM gas temperature, the collapse of gas into low-mass
halos will be suppressed throughout the Universe.\footnote{Another
possibility is metal-enrichment: once the IGM is polluted, massive
PopIII stars may stop forming, and remnant BHs will become much less
common.  However, this is less attractive, because the metals will be
much more localized around pre-existing sources (at least much more
than X-ray photons), and cannot produce a global feedback (though
clustering can make metal-feedback more effective; \citealt{Kramer+06}).}

We examine whether X-rays produced from the earliest BHs themselves
may provide sufficient heating to avoid the overproduction of the
$10^{5-7} {\rm M_\odot}$ BHs at $z\sim 6$.  Our study extends that of
TH09, by self-consistently treating global X-ray heating and the
corresponding rise in the halo mass scale for seed formation and BH
growth.

This paper is organized as follows.  In \S\ref{sec:methods}, we
describe our modeling the co-evolution of BHs
(\S\ref{subsec:BHgrowth}) and the thermal and ionization state of the
IGM (\S\ref{subsec:feedback}).  Results of the coupled evolution of BH
growth and IGM heating are presented in \S\ref{sec:results}.  In
\S\ref{sec:discuss}, we discuss earlier studies on high-redshift
miniquasar feedback, and offer general conclusions on viable models of
the $z>6$ quasar SMBHs.  Finally, we summarize our main conclusions in
\S\ref{sec:conclude}.

\section{Methods}
\label{sec:methods}

We couple a merger-tree BH assembly model with radiative-transfer
calculations of the global heating and ionization of the IGM.  Our
semi-analytic method self-consistently models the intertwined
evolution of the entire population of accreting nuclear BHs 
in halos with masses $M\ga3\times10^4\Msol$ in the redshift range
$6<z<45$,
with a statistical representation equivalent to $\approx$ 5 comoving
Gpc$^{3}$.

\subsection{Black hole growth}
\label{subsec:BHgrowth}

Following earlier work (e.g., \citealt{VHM03}; \citealt{Bromley+04};
\citealt{YooMirald04}; \citealt{VolRees06}; \citetalias{TH09}), we
compute the hierarchical growth of dark matter (DM) halos by means of Monte Carlo
merger trees in the extended Press-Schechter formalism
\citep{LaceyCole93, SomerKolatt99, Zhang+08}, and couple it with a semi-analytical model to
follow the growth and dynamics of BHs.  We employ a $\Lambda$CDM
cosmology with the parameter values
$h=0.704$,
$\Omega_{\Lambda}=0.728$, $\Omega_{\rm m}=0.272$,
$\Omega_{\rm b}=0.045$ and
$\sigma_{8}=0.81$ \citep{Jarosik+11}.
We start at $z=6$ and model the assembly history of the full halo mass
function above $M_{\rm halo}>10^{8}\Msol$ at this redshift, following
the method of \citetalias{TH09}.  The reader is referred to that paper
for a detailed description of the Monte-Carlo algorithm; here we
summarize the most important features and highlight the improvements
over \citetalias{TH09}.

The rarest, most massive halos are simulated individually, which
determines the cosmological volume represented by our suite of
simulations.  The population of less massive halos is divided into
logarithmic mass bins of width $\Delta\log(M_{\rm halo}/{\rm
M_\odot})=0.5$; within each bin, a sufficiently large number ($\sim
10^{2-3}$) of unique halos are simulated to obtain a statistical
sample for low-mass halos. The BHs in these low-mass halos are then
counted multiple times (``cloned'') to represent the same large volume
as for the most massive halos.  Our most massive bin consists of three
halos with $M_{\rm halo}>10^{12.7}\Msol$, from which we infer that our
suite of merger trees represents a statistical sample equivalent to a
comoving volume of $\sim 5~{\rm Gpc}^{3}$.  Our results are based on a
total of $\approx 10^4$ merger trees in each model.

The most significant change relative to \citetalias{TH09} is that we
have improved our halo mass resolution to correspond to a virial
temperature of $T_{\rm vir}=400$K. Above this value, the gas can
collapse through efficient cooling by ${\rm H_2}$
\citep{Haiman+96,Tegmark+97,Machacek+01}.  We follow the merger
history of each halo down to this limit.  As a result, our merger
trees also extend to higher redshift (as large as $z_{\rm
max}\approx 45$), and to lower masses
(as low as $3\times 10^4\Msol$).\footnote{
The relative streaming velocities ($1-10 \km \s^{-1}$)
between baryons and dark matter can increase
the mass threshold for halo virialization and
delay PopIII star formation (\citealt{Tseliak+11}; \citealt{Greif+11a}; 
Li, Tanaka and Haiman, in preparation.)
}

At each redshift, the comoving density $\rho_{\rm BH}(z)$ of BHs
residing in (proto-)galactic nuclei changes due to a combination of
three effects: accretion onto the holes already present at redshift
$z$, creation of new seeds in small halos, and ejections due to
gravitational recoil;
\beq
\dot{\rho}_{\rm BH, net}(z)\; = \;\dot{\rho}_{\rm acc}(z) 
\;+\; \dot{\rho}_{\rm creat}(z)\; -\; \dot{\rho}_{\rm eje}(z)\;. 
\label{Eq:rhodot}
\eeq
BH accretion deposits high-energy photons into the IGM, heating it and
affecting the conditions for subsequent BH growth.  Below, we describe
our model implementation of each of these components.

\subsubsection{Seed black hole formation}
\label{subsubsec:seeds}

Starting from the highest redshift in each merger tree, we follow all
of the branches of the tree towards lower redshifts. We place seed BHs
in halos when they first reach a virial temperature of $400 \K$. This
corresponds to a mass threshold
\beq
M_{\rm halo}^{({\rm seed})}\gta 9.1\times 10^{4}\left(\frac{\mu}{1.2}\right)^{-3/2}\left(\frac{1+z}{21}\right)^{-3/2}\Msol,
\label{eq:M400K}
\eeq
where $\mu$ is the mean molecular weight.

The initial mass function (IMF) of the first stars remains highly
uncertain.  Simulations that include the effects of radiative feedback
from the accreting protostar had suggested that the maximum mass is
$\approx 320 \Msol$ \citep{OmukaiPalla03, Ohkubo+09}, with recent
studies yielding values as low as $30 \Msol$ \citep{Stacy+12} to $43
\Msol$ \citep{Hosokawa+11}.  Recent simulations have also suggested
that hydrodynamical turbulence in star-forming halos may place {\it
typical} masses of PopIII stars significantly lower than previously
believed, perhaps $M_{*}\gta 10-50\Msol$; that their IMF was less
steep than PopI stars, and that they may have formed in small groups
or clusters \citep{Turk+09,Stacy+10,Greif+11b,Turk+12,Greif+12}.  For
our purposes, what matters is only that there is a reasonable chance
for forming {\em at least one} massive PopIII star in a minihalo that
can leave behind a seed BH.  For concreteness, we chose an IMF with
stellar masses in the range $20\Msol < M_{*} < 320 \Msol$, with a
Salpeter \citep{Salpeter55} power-law slope $dn/d\log M_{*}\propto
M^{-1.35}$.  This choice is conservative in that most of the seed BHs
have low masses -- but we emphasize that it is still possible that
most, or even all, PopIII stars had still lower masses and left no
seed BHs.

For simplicity, we also assume that {\em at most one} massive star
forms per halo, because the halo is metal-polluted once it undergoes a
star-formation episode. (This assumption is conservative: massive
stars outside the pair instability SN range are thought to collapse
directly into a BH without ejecting metals; \citealt{Heger+03}.)
Pristine halos that merge with such halos are also considered to be
polluted.  We do not allow seed BH formation in metal-polluted halos;
we consider such halos to be unable to form massive stars whose BH
remnants have masses of more than several $\Msol$ \citep{Heger+03}.

The seed BH masses are modeled using fitting formulae to the
simulation results of \cite{ZhangW+08}, who calculated the remnant
masses of metal-free stars undergoing core-collapse explosions while
accounting for the mass of the ejecta that falls back onto the
remnant.  Specifically, we prescribe
\beq
M_{\rm BH}=
\left\{ \begin{array}{ll}
       \frac {3}{4} (M_{*}-20\Msol) +2\Msol 
         & \mbox{if $M_{*}\le 45\Msol$}\\
      \frac {5}{12} (M_{*}-20\Msol)
          & \mbox{if $M_{*}>45\Msol$}\end{array} \right.
\label{eq:IMF}
\eeq
and assume that stars in the pair-instability mass range
$140\Msol<M_{*}<260\Msol$ leave no BH remnant.  

For reference, we note that the above differs from the earlier
treatment of \citetalias{TH09}, in which all of the BH seeds were
assumed to have the same mass of $100\Msol$.
In the present model, the mean PopIII stellar mass is
$\langle M_{*}\rangle \approx 49\Msol$,
and the mean seed BH mass is $\langle M_{\rm BH}\rangle \approx13 \Msol$.

\subsubsection{Black hole accretion}
\label{subsubsec:accretion}

As our fiducial accretion model, we take a simple scenario in which
all seed BHs grow continuously (at least in a time-averaged sense) at
a fraction $f_{\rm Edd}$ of the Eddington accretion rate,
\beq
\dot{M}=f_{\rm Edd}\dot{M}_{\rm Edd};
\label{eq:expon}
\eeq
\beq
\dot{M}_{\rm Edd}
=2.2\times 10^{-3} \frac{1-\epsilon}{\epsilon}M_{\rm BH} \Myr^{-1}\propto M_{\rm BH}.
\label{eq:Edd}
\eeq
Here $\epsilon$ is the radiative efficiency, which we take to be 0.07
\citep[cf.][]{MerloniHeinz08, Shankar+09a}.  With this prescription
all BHs grow exponentially with an e-folding timescale of $34 f_{\rm
Edd}^{-1}\Myr$.
We also consider a more realistic prescription in which the accretion
rate is further limited by the Bondi rate $\dot{M}_{\rm Bondi}$, i.e.
\beq
\dot{M}={\rm min}(f_{\rm Edd}\dot{M}_{\rm Edd},\dot{M}_{\rm Bondi});
\label{eq:Bondilim}
\eeq
\beq
\dot{M}_{\rm Bondi}=4\pi G^{2}M_{\rm BH}^{2}\frac{\rho_{\rm gas}}{c_{\rm s}^{3}}
\propto M_{\rm BH}^{2}.
\label{eq:Bondi}
\eeq
Here $\rho_{\rm gas}$ and $c_{\rm s}$ are the gas density and sound
speed, respectively.  In general, in the cold and dense cores of the
cuspy gas density profiles of minihalos, in which gas cools
efficiently \citep{Abel+02,Bromm+02}, the Bondi rate exceeds the
Eddington rate, and so does not impose a limitation
(\citealt{Turner91}; \citetalias{TH09}).  On the other hand, the
growing seed BHs can find themselves in a much lower-density medium
for several reasons. First, the progenitor star could have blown most
of the gas out of the host halo, so that the seed BH is born in a
low-density gas \citep{Whalen+04,Alvarez+09}.  Second, although the
photo-heated gas could cool (via Compton scattering on the CMB) and
fall back into the halo, it will retain excess entropy and may not be
able to form ${\rm H_2}$ and cool again, especially if a strong
Lyman-Werner background is present \citep{Haiman+97,OH03}.  In this case,
the gas will be in a pressure-supported configuration at much lower
density (e.g. \citealt{Shapiro+99,Ricotti09}).\footnote{Note that
progenitor stars with masses above $40~{\rm M_\odot}$ are expected to
collapse directly into a BH and not eject any metals \citep{Heger+03},
eliminating the option of cooling via heavier elements.}  Third, even
as the halo undergoes mergers with other minihalos, the gas in those
partner halos may have been unable to cool, as well, owing to the
presence of a Lyman-Werner background.  Fourth, the seed BH can be
kicked out via gravitational recoil from its parent halo, and spend a
prolonged period wandering in the lower-density halo outskirts
(\citealt{BlechaLoeb08}; \citetalias{TH09}; \citealt{Guedes+09}).

Rather than attempting to model the above effects self-consistently,
we simply adopt a fiducial Bondi rate, scaled to the central gas
density of the host halo.  In an NFW DM halo \citep{NFW96}, the gas
density profile has a central core that is proportional to the
background IGM gas density \citep{Madau+01}, $\rho_{\rm gas}^{\rm
(core)}\propto \rho_{\rm gas}^{\rm (IGM)}\propto (1+z)^{3}$.
Following \cite{Ricotti09}, we assume an isothermal central gas
equation of state and adopt the functional form
\beq
\delta_{\rm core}= \frac{\rho_{\rm gas}^{\rm (core)}}{\rho_{\rm gas}^{\rm (IGM)}}-1
\approx \Delta_{\rm gas}
\frac{\left(1-0.2\beta/3\right)(c/0.2)^{3}}
{(c/0.2)^{3(1-0.2\beta/3)}-0.2\beta/3}.
\label{eq:rhocore}
\eeq
Above, $\Delta_{\rm gas}\approx 5$ is the mean gas overdensity of the
halo; $c$ is the halo concentration parameter; $\beta\equiv
c\Gamma/f(c)$ is defined in terms of the parameter $\Gamma=T_{\rm
vir}/T_{\rm IGM}$ and $f(c)=\ln(1+c)-c/(1+c)$.  An important feature
of this model is that the core density depends on the IGM temperature
(this dependence arises through the external pressure from the IGM
near the halo's virial radius).  In the limit of large $\beta\propto
T_{\rm vir}/T_{\rm IGM}$, $\delta_{\rm core}\approx 600 c^{3}$.
For the concentration parameter, we adopt
\beq
c=4.5\left(\frac{M_{\rm halo}}{10^{6}\Msol}\right)^{-1/8}\left(\frac{1+z}{21}\right)^{-1},
\label{eq:conc}
\eeq 
where the normalization and the dependences on $M_{\rm halo}$ and
$z$ are based on the simulation results of \cite{Bullock+01} and
\cite{Wechsler+02}.  Several studies have found that the redshift
dependence of the concentration parameter weakens at high redshifts
$z\ga 4$, and that $c$ has a characteristic minimum value $c_{\rm
min}\sim 3-5$ \citep{ZhaoD+03, ZhaoD+09, Klypin+11}.
We impose a conservative lower limit $c\ge 3$ to
equation \ref{eq:conc}; this still allows for the Bondi accretion rate
of our BHs to be many orders of magnitude below the Eddington rate.

To summarize, the core gas density is a function of (i) the mean IGM
gas density (or redshift), (ii) the halo concentration parameter $c$,
and (iii) the parameter $\Gamma=T_{\rm vir}/T_{\rm IGM}$.  Given the
large uncertainties in modeling the relevant gas density profiles,
this Bondi accretion model can be more generally understood in terms
of the core overdensity $\delta_{\rm core}$.  The critical value of
$\delta_{\rm core}$ at which $\dot{M}_{\rm Bondi}=f_{\rm
Edd}\dot{M}_{\rm Edd}$ is given by
\beq
\frac{\delta_{\rm crit}}{f_{\rm Edd}}\approx 2\times 10^{5}
\left(\frac{M_{\rm BH}}{50\Msol}\right)^{-1}
\left(\frac{T_{\rm vir}}{400\K} \right)^{3/2}
\left(\frac{1+z}{21}\right)^{-3}.
\label{eq:deltacrit}
\eeq

\subsubsection{Black hole mergers and gravitational recoil}
\label{subsubsec:recoil}
Following the merger of two DM halos that each contain a central BH,
we assume that the BHs promptly form a binary and coalesce, aided by
the gas surrounding them (e.g. \citealt{Escala+04,Cuadra+09};
see \citealt{KulkarniLoeb12} for the possible consequences on
the high-redshift SMBH population if merger timescales
for SMBHs are longer than those for their host halos).
A recoil velocity is assigned to the merged binary, with a random
value drawn from the fitting formulae given by \cite{Lousto+10},
assuming a uniform distribution for the spin magnitudes (in the range
$0<a<0.93$) and angles with respect to the binary's orbital plane
($0<\theta<\pi/6$).  The latter choice is motivated by
\citet{Bogdanovic+07} and \citet{Dotti+09}, who have argued that
coplanar circumbinary disks can cause moderate spin alignment.

Radial trajectories of the kicked BHs are computed as in
\citetalias{TH09}, including the effects of dynamical friction and
assuming an NFW DM potential superimposed with an isothermal gaseous
profile ($\rho_{\rm gas}\propto r^{-2}$).  The latter is taken to be
an isothermal cusp in models with an exponential accretion
prescription (eq. \ref{eq:Edd}), and the gas profile with a core
(eq. \ref{eq:rhocore}) for models with the Bondi-limited prescription
(eq. \ref{eq:Bondi}).  Instead of computing the radial trajectories of
recoil explicitly for each merger event, we compare the recoil
velocity against a retention velocity threshold $v_{\rm ret}(z,M_{\rm
halo}, M_{\rm BH})$ which is interpolated from a precompiled table.
We consider a recoiling BH to be ejected if it does not return to the
halo center on a radial orbit within a Hubble time.

In general, we have found that, somewhat counter-intuitively, the
final SMBH mass function at $z\gta 6$ is not highly sensitive to the
precise spin prescription (\citetalias{TH09}), or to the central mass
distribution of the host galaxy.  This is because the recoil velocity
is a strongly increasing function of the mass ratio $M_{2}/M_{1}\le
1$, and the effects of gravitational recoil may be understood rather
simply in terms of critical {\it mass ratios}.  Nearly-equal mass BH
binaries should almost always be ejected, barring an ultra-massive
host halo, strong gas drag from an extremely dense central gas
profile, or highly specific, recoil-minimizing BH spin
orientations \citep{Haiman04}.  Conversely, mergers with mass ratios
$\ltsim 1:20$ result in kicks of only $\ltsim 20-30\km\s^{-1}$
regardless of BH spins, and cannot eject BHs from deep inside their
halos.  Fully-grown SMBHs are ejected only very rarely, as only the
most extreme mass ratios and spins can produce kicks large enough to
remove them from their massive host halos.  Therefore, the spins and
the host gravitational potential can influence only the fates of BHs
that merge with mass ratios $\sim 1:10$ in low-mass halos early on.

In our models, the ancestors of $z\sim 6$ SMBHs tend to be BHs that
were sufficiently more massive than their merger partners, and whose
mergers occurred in sufficiently massive host halos.  The probability
that particularly massive BHs survive gravitational recoil events
increases with time, as they become successively more massive with
respect to their contemporaries and the gravitational potential of
their host halos grow deeper (\citealt{VolRees06}; \citetalias{TH09}).

\subsection{IGM heating and radiative feedback}
\label{subsec:feedback}

\subsubsection{(Mini-) quasar emission spectra}

We model all accreting black holes as light-bulb (mini-) quasars with
Eddington luminosities that are ``on'' for a fraction
$\dot{M}/\dot{M}_{\rm Edd}$ of the time, and ``off'' the rest of the
time.  We assume that a fraction $f_{\rm cor}=0.1$ of the total energy
is emitted by a hot X-ray corona, modeled as a power-law spectrum
above 1 keV with a photon index $\Gamma=2.0$.  The rest of the
emission is modeled as a multicolor Shakura-Sunyaev accretion disk
\citep{SS73}, with $\dot{M}=\dot{M}_{\rm Edd}$ and viscosity parameter
$\alpha=0.1$.  We model the spectrum as a one-zone graybody
photosphere \citep{Blaes04} using the fitting formulae of \cite{TM10}
for the graybody source function; the disk emission is scaled down by
a factor $1-f_{\rm cor}$ at all frequencies to accommodate the coronal
emission.  All miniquasar photons with energies above $13.6$ eV are
assumed to escape the host halo \citep{Whalen+04}.

\begin{figure}
\includegraphics[scale=0.44,angle=0]{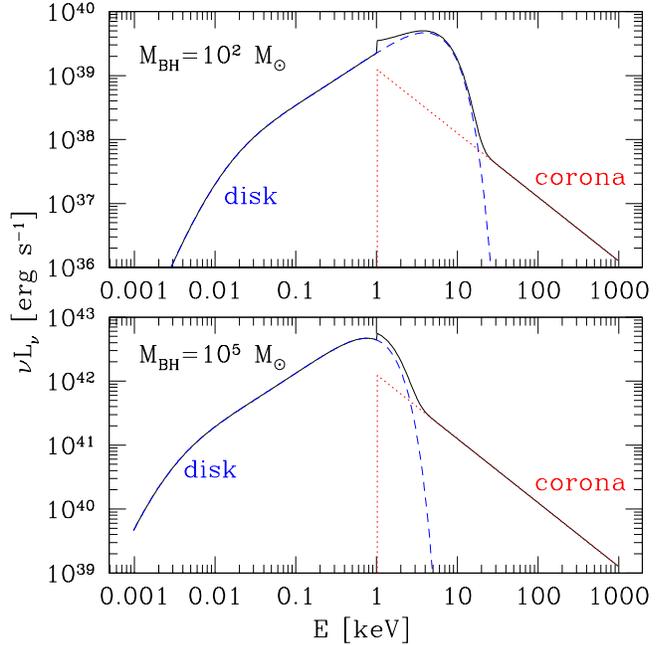}
\vspace{-0.2in}
\caption{Model spectra of accreting BHs, assumed to be light bulbs
  radiating at their Eddington luminosity.  {\it Top:} $M_{\rm
    BH}=10^{2}\Msol$.  {\it Bottom:} $M_{\rm BH}=10^{5}\Msol$.  The
  total spectrum (solid black) consists of an $\alpha$-viscosity
  accretion disk (dashed blue) and an additional power-law corona at
  $>1$keV (dotted red).  }
\label{fig:spectra}
\end{figure}

Figure \ref{fig:spectra} shows two examples of our model spectra, for
miniquasars with $M_{\rm BH}=100\Msol$ and $10^{5}\Msol$. For the
entire mass range of interest, miniquasars have significant emission
above $1$ keV (analogous to high-mass X-ray binaries and ultraluminous
X-ray sources).  Our qualitative findings are robust as long as
miniquasars are abundant in the early Universe and are prolific
producers of $\sim 1$keV X-ray emission \citep[cf.][]{Venkat+01,
  Madau+04, RicOst04}.

\subsubsection{IGM heating and cooling}

We treat the IGM as a primordial H+He gas with a mass fraction $Y_{\rm
  He}=0.24$ in He.  Initially, it cools adiabatically as
\beq
T_{\rm IGM}(z)=2.73 (1+z_{\rm t})\left(\frac{1+z}{1+z_{\rm t}}\right)^{2}\K \approx 9 \left(\frac{1+z}{21}\right)^{2}\K, 
\label{eq:TIGMad}
\eeq
where $z_{\rm t}\approx 140$ is the redshift at which the IGM and CMB
temperatures decouple \citep[e.g.,][]{Peebles93}.

We follow the temperature and ionization state of the IGM from high to
low redshift, together with the evolving BH population, using small
timesteps with $\Delta z\approx 0.1$.  (At $z>30$, where miniquasar
heating is negligible, we use $\Delta z=1$.)  Across each timestep,
the high-energy photons from the accreting BHs heat and ionize the
IGM; the conditions in the IGM in turn influence the formation and
growth of seed BHs.  The ionizing photon background is built up via
the cumulative miniquasar emission, with the comoving specific
luminosity density
\beq
\epsilon_{\nu}(z)=\frac{\sum_{\rm BHs}L_{\nu}(z)}{V},
\eeq
where $V=5~{\rm Gpc}^{3}$ is the comoving volume of our merger-tree
simulations.  To compute the mean background flux at redshift $z$, we
solve the cosmological radiative transfer equation \citep{HaardtMadau96}
\beq
j_\nu(z) =\frac{1}{4\pi} \int_z^{\infty}\; \epsilon_{\nu}(z^{\prime})
\,e^{-\tau_\nu(z,\, z^{\prime})}\,
\frac{(1+z)^3}{(1+z^\prime)^3}\,c\frac{dt}{dz^{\prime}}\; dz^{\prime},
\label{Eq:flux}
\eeq 
where 
\beq
\tau_\nu(z,\,z^{\prime})\;=\;\int_z^{\prime}
\Sigma_j\,\sigma_\nu^j \,n_0^j
\,(1+z^{\prime\prime})^3\,
c\frac{dt}{dz^{\prime\prime}}\,dz^{\prime\prime}
\label{Eq:tau}
\eeq
is the opacity.  Above, $n_0^j$ denotes the comoving number density of
ion species $j$ (HI, HeI, or HeII), $\sigma_\nu^j$ is the
photoionization cross section, and
$\nu\equiv\nu^{\prime\prime}(1+z)/(1+z^{\prime\prime})$, where
$\nu^{\prime\prime}$ is the emission frequency at redshift
$z^{\prime\prime}$.

With the ionization background given by Eq.(\ref{Eq:flux}), we follow
the thermal evolution of the medium by solving, for each time-step
$dt$, the energy conservation equation:
\beq
\frac{d u}{dt}=-p \frac{d}{dt} \left(\frac{1}{\rho_{\rm gas}}\right)
- \frac{\Lambda_{\rm net}}{\rho_{\rm gas}},
\label{Eq:energy}
\eeq 
where $u$, $p$ and $\rho_{\rm gas}$ are the specific internal energy,
pressure and density of the gas, respectively, and $\Lambda_{\rm net}$
is the net heating/cooling rate per unit volume.  Heating includes
Compton and photoionization heating, while cooling includes line and
continuum cooling (and Compton cooling once the IGM temperature rises
above $T_{\rm CMB}$); the first term on the right hand side represents
cooling due to adiabatic expansion. The energy equation is solved
coupled with the chemistry equations, solving for the abundances of H,
H$^+$, He, He$^+$ and He$^{++}$.  The physics of heating and
ionization by soft X-ray photons has been described in a number of
works (\citealt{HaardtMadau96}; \citealt{Venkat+01}; \citealt{FS10}).
X-rays propagate to much larger distances than UV photons before being
absorbed, photo-ionizing He and H atoms.  These photoionizations
produce fast photo-electrons,which then partially photoionize the gas through
repeated secondary ionizations. We use the recent results by
\cite{FS10} for the fraction of the energy of each fast photo-electron
that is used for ionizations and heating. Specifically, we use simple
fitting functions to the fractions shown in their Figure 4, and we
neglect the dependence on the photoelectron's energy, which is weak
for the relevant range considered here ($E\gta 1$keV).

\section{Results: Effect of IGM heating on BH growth}
\label{sec:results}

\subsection{SMBH assembly without feedback}

\begin{figure}
\includegraphics[scale=0.44,angle=0]{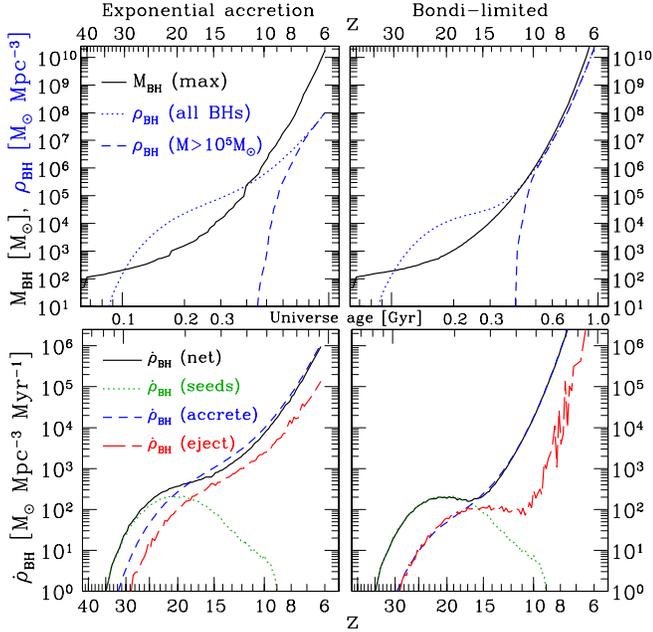}
\vspace{-0.25in}
\caption{The evolution of the BH population in two models with
  different accretion prescriptions without global feedback.  {\it
    Left panels:} All BHs accrete exponentially at a fraction $f_{\rm
    Edd}=0.55$ of the Eddington limit.  {\it Right panels:} Accretion
  is limited by the local Bondi rate (see text) and by $f_{\rm
    Edd}=0.8$.  {\it Top panels: } The mass of the most massive BH in
  the simulation (solid black curves), and the comoving density of all BHs
  (dotted blue) and of only BHs with $M>10^{5}\Msol$ (dashed blue).
  {\it Bottom panels:} The growth rate $\dot{\rho}_{\rm BH}$ of the
  global BH mass density (solid black), and contributions to it from
  seed creation (dotted green), gas accretion (short-dashed blue) and
  recoil-ejections (long-dashed red).}
\label{fig:nofb}
\end{figure}

To illustrate the overproduction problem, we first show in
Figure~\ref{fig:nofb} the SMBH growth history in two reference models
without global feedback.  The left-hand panels show the results of a
model assuming exponential accretion (eq. \ref{eq:expon}) with $f_{\rm
Edd}=0.55$; the right panels show a second reference model assuming
Bondi-limited accretion (eq. \ref{eq:Bondilim}) and $f_{\rm Edd}=0.8$.
In both cases, the BHs do not ``know'' about the temperature evolution
of the IGM. The core gas densities in the Bondi model are computed
through eq. \ref{eq:rhocore} as if the IGM cooled adiabatically.

The top panels show the growth of the mass of the most massive BH in
the simulation (solid black curves), along with the comoving mass
density of all BHs (blue dotted curves) and of only massive BHs with
$M>10^{5}\Msol$ (blue dashed curves).  The most massive BH is a
particularly massive seed with $M_{\rm BH}\sim 100 \Msol$, left behind
by a progenitor star with $M_{*}\sim 260 \Msol$, just above the pair
instability window.  Because the most massive SMBHs in our models are
products of multiple mergers, their masses scale roughly linearly with
the maximum allowed seed mass.

Given that the local SMBH mass density is $\rho_{\rm BH,0}\sim 4\times
10^{5}\Msol \Mpc^{-3}$ \citep{Marconi+04, Shankar+04} and that
$\gta 90\%$ of this mass density appears to have accreted at $z<6$
during luminous quasar phases \citep[e.g.,][]{Shankar+09a}, the mass
density of massive BHs with $M\gta 10^{5-6}\Msol$ at $z\sim 6$
should be at most $\sim {\rm few}\times10^{4}\Msol \Mpc^{-3}$.  Both
models presented in Figure~\ref{fig:nofb} overpredict the global SMBH
mass density by several orders of magnitude, due to numerous seed BHs
growing exponentially near the Eddington rate.  

The failure of the Bondi-limited accretion model to avoid this
overproduction is not obvious, and is worth explaining in some detail.
We find that the seeds surviving their recoil events and producing the
$>10^9~\Msol$ SMBHs at $z=6$ are those with initial masses of $\sim
30-100\Msol$.  As can be seen from equations \ref{eq:rhocore} and
\ref{eq:deltacrit}, these seeds, born in halos with $T_{\rm vir}\ga
400$K, quickly get out of their Bondi-phase and switch to the
Eddington-limited regime, provided that the central gas density in the
halo is sufficiently large (i.e. that the halos have virial
temperatures sufficiently above the IGM temperature threshold).
As a result, the most massive BHs in the Bondi-limited and the
exponential models are similar (compare the black curves in the top
left vs. top right panel in Figure~\ref{fig:nofb}).  As equation
\ref{eq:deltacrit} shows, the Bondi-limit is important for lower-mass
seeds, and it also becomes more important at lower redshifts, where
the characteristic densities are lower.  This is indeed causes a
``dip'' and slows down the overall growth of the BH population by
accretion in the range $10\ltsim z\ltsim 20$.  In this range,
$\rho_{\rm BH}$ is dominated by relatively low--mass ($\ltsim
10^3~\Msol$) BHs, which arise from relatively low--mass seeds, and
which suffer prolonged episodes of Bondi-limited accretion.  However,
as Figure~\ref{fig:nofb} shows, this slow-down is insufficient: the
total mass density in $\gta 10^5~\Msol$ BHs by $z\ltsim10$ is
dominated by relatively massive and early-forming seeds, which did not
spend a significant time in a Bondi-limited stage.

The bottom panels of Figure \ref{fig:nofb} show the growth rate
$\dot{\rho}_{\rm BH}$ of the comoving BH mass density (solid black
lines), and its three components: creation of new seed BHs (dotted
green lines), accretion onto existing seeds (short-dashed blue lines),
and ejections due to gravitational recoil (long-dashed red lines).
The growth of the total BH density can be broadly divided into two
distinct epochs, being dominated by the formation of new seeds at
early times ($z\gta 20$) and by accretion at late times ($z\ltsim15$).
The seed formation history is identical in the two models
(\S\ref{subsubsec:seeds}).  In the Bondi-limited case,
only BHs with $M_{\rm BH}$ exceeding a critical value $M_{\rm crit}$
(see \ref{eq:deltacrit})
grow exponentially and the rest accrete at a much lower rate.  This
leads to a wider distribution in BH masses at late times, which
results in fewer BH mergers with comparable masses and thus fewer
ejections by gravitational recoil.  In the exponential growth model,
BHs tend to have similar growth histories; two BHs that have similar
masses at a given time will also have similar masses at a later time,
unless one grows much more rapidly via mergers. Thus, pairings of BHs
of comparable masses are common, resulting in much higher rates of BH
ejection.  In both models, and generically for assembly models of this
type, ejected BHs tend to be somewhat lower-mass BHs
residing in lower-mass halos (\S\ref{subsubsec:recoil}).
As seen in the bottom left panel in Figure~\ref{fig:nofb}, the global
mass density of nuclear BHs can temporarily {\it decrease} during
phases in which the ejections of low-mass BHs outweigh the sum of
accretion onto high-mass BHs and the seed formation rate.

In both accretion models, the growth rate of the overall population
$\dot{\rho}_{\rm BH}$ becomes unrealistically high by $z\sim
9$.\footnote{We have checked that despite these high rates, the $z>6$
  SMBHs in our models never consume more than 10 percent of the
  baryons in a halo. The most massive SMBHs typically have less than
  one percent of the host halo's baryon mass.
  }
For reference, to avoid overproduction, this rate should have a
time-averaged value of
\beq
\langle \dot{\rho}_{\rm BH}(z>6)\rangle
\ltsim  \frac{0.1\times \rho_{\rm BH, 0}}{\Delta t (z>6)}
\sim 60 \Msol\Mpc^{-3}\Myr^{-1}.
\label{eq:rhodotmax}
\eeq

\subsection{SMBH assembly with global IGM warming}
\label{subsec:withfb}
The most conspicuous consequence of IGM heating is the inhibition of
the gravitational collapse of gas into low-mass halos.  The
cosmological Jeans mass \citep[e.g.,][]{BarkanaLoeb01}
\begin{align}
\label{eq:MJ}
M_{\rm J}(z) &= \left( \frac{5kT_{\rm IGM}}{G\,\mu\, m_p}\right)^{3/2}\,
\left(\frac{3}{4\pi\rho_{\rm b}} \right)^{1/2}\\
&\approx 6.8\times 10^{5}
\left(\frac{\mu}{1.2}\right)^{-3/2}
\left(\frac{T_{\rm IGM}}{100\K}\right)^{3/2}
\left(\frac{1+z}{21}\right)^{-3/2}
\Msol\nonumber
\end{align}
will rise, and gas inside halos whose masses are below this threshold
will not collapse to form stars and seed BHs.  It has been shown
\citep{Gnedin00} that the so-called filtering mass, which depends on
the temperature history of the IGM, provides a better fit to gas
fractions in low-mass halos in numerical simulations than the
instantaneous Jeans mass.  The filtering mass lags the evolution of
the Jeans mass, and is therefore {\em lower} than the Jeans mass,
during the period when the IGM temperature is rising. However, we note
that the definition in equation~\ref{eq:MJ} is a factor of 8 lower
than the one adopted by \cite{Gnedin00}, and is, in fact, very close
to the filtering mass shown in his Figure 3 at the relevant redshifts
($z\gta 10$).\footnote{\cite{NaozBark07} showed that the gas density
and pressure gradients in the initial conditions after recombination
decreases the filtering mass. However, this is below our minimum halo
mass imposed by cooling, and does not affect our results.}  We note
that the impact of IGM heating may extend beyond the fiducial Jeans
mass: even in a halo above the Jeans mass, gas falls in, at least
initially, along a higher adiabat from the pre-heated IGM---this
could prevent the gas from condensing to a sufficiently high enough
density to activate ${\rm H_2}$ cooling.  \cite{MBA03} and \cite{KM05}
both found evidence in 3D simulations that heating the IGM by X-rays
can suppress baryonic infall into low-mass halos.

Another consequence of IGM heating may be to suppress the BH accretion
rate in pre-existing halos.  \cite{MBA03} and \cite{KM05} found that
an early soft X-ray background can elevate the mean gas temperature
inside low-mass minihalos with masses below the collapse threshold.
(However, both studies found that the X-ray feedback on cold gas may
be {\it positive} in more massive minihalos; see
\S\ref{subsec:compare} below.)  As the IGM heats, the inflow of cool
gas into existing halos whose masses fall below the collapse threshold
would be suppressed. The heating and evaporation of small halos could
in turn impact accretion onto BHs residing in more massive halos, by
reducing the amount of cold and dense gas supplied by minor mergers.
Furthermore, the accumulating X-ray background can heat the gas inside
low-mass halos near or above their virial temperatures, reducing the
gas density, or photoevaporating the halo entirely \citep{Shapiro+04,
AhnShapiro07}, shutting down Bondi accretion onto the nuclear BH.

There is some circumstantial evidence that the BH accretion rate is
tied to the mass supply rate onto the parent halo.  Massive BHs appear
to have been much more active in the early Universe: most $z\sim 6$
quasars appear to be shining near or slightly above their Eddington
luminosities \citep{Willott+10b}, compared to typical Eddington ratios
of $\sim 0.1$ at $z<4$
\citep{Kollmeier+06,Kelly+10}. \citet{Shankar+10} concluded that most
SMBHs are active as quasars at $z\sim 6$, compared to less than $1\%$
at $z<2$ \citep{Shankar+09a}.  A plausible physical reason behind such
hyperactivity is the rapid mass supply to their host halos via mergers
with other halos and by accretion; \cite{Angulo+12} observed that
numerically simulated host halos of $z\sim 6$ quasars doubled in mass
in the preceding $100$ Myr.  

Motivated by these considerations, we simulate two models of SMBH
assembly that include negative feedback due to IGM heating.  In both
models, the cosmological Jeans mass rises due to IGM heating by
miniquasars; new seed BHs are only formed in chemically pristine halos
with masses above the Jeans mass.  In the first {\em
(``exponential'')} model, we assume that BHs grow exponentially with
$f_{\rm Edd}=0.55$ as in one of the reference models, but that this
growth is enabled by the supply of cold gas from the merger-driven
growth of the host halo.  Thus, we halt BH accretion one e-folding
time ($\approx 62 \Myr$, comparable to a typical quasar lifetime;
e.g. \citealt{Martini04}) after the most recent merger of their host
halo with another halo. We additionally require that the merger
partner halo exceeds the Jeans mass (so that it contains cold gas).
In the second {\em (``Bondi'')} model, we assume that BH accretion is
instead limited by the Bondi rate in the halo nucleus, which we model
as in the second reference model. We again take $f_{\rm Edd}=0.8$ and
$\rho_{\rm gas}=\rho_{\rm min}$.  However, the central gas density of
the halo depends on the ratio $\Gamma=T_{\rm vir}/T_{\rm IGM}$
(eq. \ref{eq:rhocore}) -- as the miniquasar emission heats the IGM,
the core densities in low-mass halos decrease and the Bondi accretion
rates of their BHs are suppressed.

\subsubsection{Global IGM warming: results}
\label{subsubsec:warmingresults}

\begin{figure}
\includegraphics[scale=0.44,angle=0]{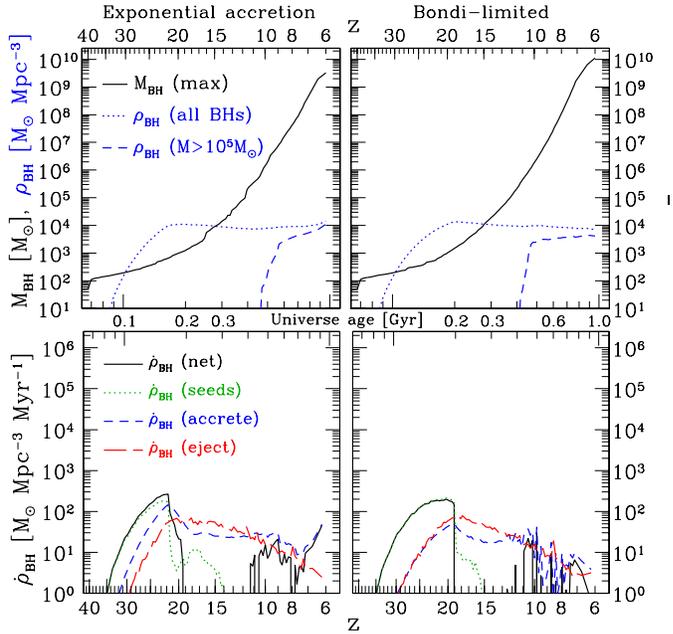}
\vspace{-0.25in}
\caption{Same as Figure~\ref{fig:nofb}, but global IGM heating by
accreting BHs suppress BH formation and growth at late times.
{\it Left panels:}
BHs accrete exponentially as in the left panels of
Figure~\ref{fig:nofb}, but accretion is halted if host halos do not
continue merging with massive halos containing cold, collapsed gas.
{\it Right panels:}
BH accretion is limited by the local Bondi rate as in the
right panels of Figure~\ref{fig:nofb}; the rising IGM temperature
lowers the halo gas density (eq. \ref{eq:rhocore}).  In both models,
halos below the Jeans mass are disqualified from forming new seed BHs.
}
\label{fig:withfb}
\end{figure}

Our results in both the Exponential and the Bondi models are shown in
Figure \ref{fig:withfb}.  As in Figure~\ref{fig:nofb}, the left panels
show the exponential accretion model, while the right panels show the
Bondi-limited model.  The top panels show the growth of the most
massive BH in each simulation, along with the global BH mass density
$\rho_{\rm BH}$, while the bottom panels show $\dot{\rho}_{\rm BH}$
and its components due to seed formation, gas accretion and
recoil-ejections.  The color and line style schemes are identical to
those in Figure~\ref{fig:nofb}.  For ease of comparison with the
reference models, the vertical scales of the graphs are the same as in
Figure~\ref{fig:nofb}, despite much lower values of $\dot{\rho}_{\rm
BH}$ in the models with feedback.

\begin{figure}
\includegraphics[scale=0.44,angle=0]{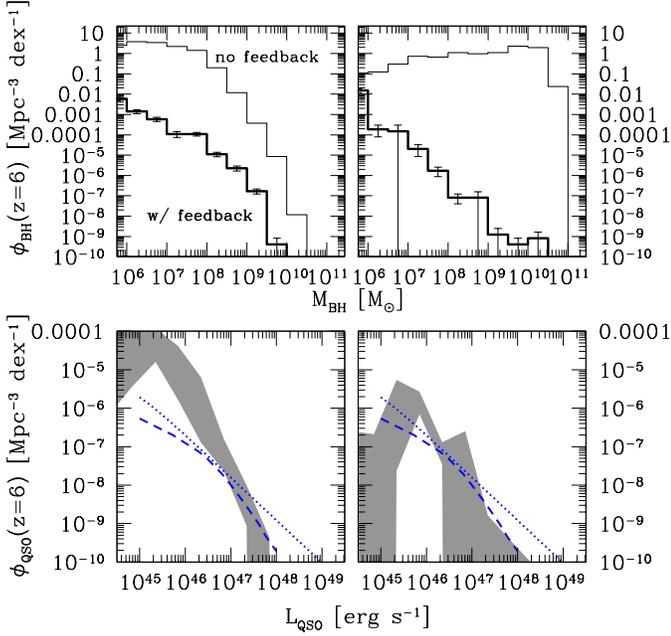}
\vspace{-0.25in}
\caption{ {\it Top panels:} The $z=6$ BH mass function in our SMBH
  assembly models.  The thin histograms show the models without
  feedback, severely overproducing the global SMBH mass density; the
  thick histograms show the models with feedback.  The error bars
  demarcate the Poisson errors from our simulated sample.  {\it Bottom
  panels:} The $z=6$ quasar luminosity function in our models with
  feedback, shaded in grey between the upper and lower limits for the
  luminosity and duty cycles of individual BHs (see text).  For
  reference, we show the obscuration-corrected luminosity function
  found by Willott et al. (2010; blue dashed curve) and the best-fit
  model for $z=6$ of Hopkins et al. (2007; blue dotted curve).  As
  with Figs. \ref{fig:nofb} and \ref{fig:withfb}, the left panels show
  exponential accretion models with $f_{\rm Edd}=0.55$, and the right
  panels show models where accretion is capped by the Bondi rate and
  by $f_{\rm Edd}=0.8$.  }
\label{fig:bhmf}
\end{figure}

In the top panels of Figure \ref{fig:bhmf}, we show the $z=6$ BH mass
function produced by our models, without (thin lines) and with
feedback (thick lines).  As with the previous two figures, left panels
show the exponential-accretion models and the right panels show the
Bondi-limited models.  Feedback clearly depresses the late growth of
$M_{\rm BH}\sim 10^{5-7}\Msol$ BHs, while leaving the most massive
objects nearly unaffected.  Note that the mass function is very steep
in the models without feedback.
This is because most seeds form within a relatively
brief cosmological epoch, and a large number of BHs
are allowed to grow at the identical exponential rate.

In the bottom panels of Figure~\ref{fig:bhmf}, we estimate the
luminosity function for the models with feedback that avoid SMBH
overproduction.  Recall that in order to treat the X-ray emission from
accreting BHs, we have approximated them as Eddington-luminosity
lightbulbs with a duty cycle $\dot{M}/\dot{M}_{\rm Edd}$.  Using this
prescription to translate the mass function into a quasar luminosity
function, however, will tend to overestimate quasar luminosities and
underestimate their counts (duty cycles).  We therefore also consider
the opposite extreme estimate for the luminosity function, in which
BHs shine at a bolometric luminosity $(\dot{M}/\dot{M}_{\rm
Edd})L_{\rm Edd}$ and have a duty cycle of $1$.  The true luminosity
function should lie within the bounds set by these two formulations,
represented in Figure \ref{fig:bhmf} by the shaded grey region.  For
reference, we plot the observational determination of the $z\sim 6$
quasar luminosity function (corrected for obscuration) of
\citeauthor{Willott+10b} (\citeyear{Willott+10b}; blue dotted lines),
as well as the fitting formula suggested by \citeauthor{Hopkins+07b}
(\citeyear{Hopkins+07b}; blue dashed lines).

The luminosity functions predicted by our
models, in particular the ``exponential'' model, may still be too steep
below $L\sim 10^{46.5}\erg\s^{-1}$ compared to the best-fit luminosity function
of \cite{Willott+10b}.
However, the faint end of the $z\approx 6$ quasar luminosity function 
remains highly uncertain.
The \citeauthor{Willott+10b} fit includes a single binned data point below
$2\times 10^{46}\erg\s^{-1}$,
and assumes that the fraction of obscured quasars (which is larger
for low-luminosity objects; \citealt{Lawrence91}) is the same as
the one found in the local Universe.
The latter point is motivated by the fact that the obscured fraction
does not appear to evolve between $z=0$ and $2$ \citep{Ueda+03},
but there has been no direct determination of the obscured fraction at $z\approx 6$.

As Figures \ref{fig:withfb} and \ref{fig:bhmf} show, both of these
models avoid the overproduction of the lower-mass ($M\sim 10^{5-7}$)
BHs, while still producing $M\gta 10^{9}\Msol$ by $z\approx 6.4$.
Note that the exponential growth model, in which the accretion rate is
capped at half of the Eddington limit, falls somewhat short of
explaining the $z=7.08$ quasar SMBH with $M\approx 2\times10^{9}\Msol$
observed by \cite{Mortlock+11} despite producing a good fit to the
$z\approx 6$ mass function of \cite{Willott+10b}.  At $z=7.08$, in a
simulation volume roughly $1/3$ of the $\sim 15$ comoving Gpc$^{3}$
covered by the UKIDSS, the most massive SMBH in this model is a factor
$\approx 4$ short.  The $z=7.08$ SMBH can be reproduced with only a
slightly higher accretion rate, $f_{\rm Edd}=0.6$; however, this
overshoots the global mass function and $\rho_{\rm BH}$ by roughly an
additional order of magnitude.  This illustrates the fact that this
SMBH is yet more difficult to accommodate in simple SMBH assembly
models. One obvious possibility that the \citeauthor{Mortlock+11} SMBH
is a mild outlier in its growth history.  While $f_{\rm Edd}=0.55$ is a
suitable mean value in the context of this model, realistically BHs
have a distribution of Eddington ratios and not a single universal
value.

In summary, these results of our illustrative models suggest a
generic, self-consistent scenario of black-hole-made ``climate
change'', in which the earliest and most massive accreting BHs heat
the IGM and suppress the formation and growth of subsequent
generations of BHs.  As in the case of global warming on Earth, the
negative effects are felt by the next generations: the BHs originally
responsible for this feedback are largely unaffected by it, because
they reside in the most massive halos that merge frequently with other
massive halos and have high central gas densities to facilitate BH
growth.

\subsubsection{IGM global warming and reionization}

\begin{figure}
\includegraphics[scale=0.44,angle=0]{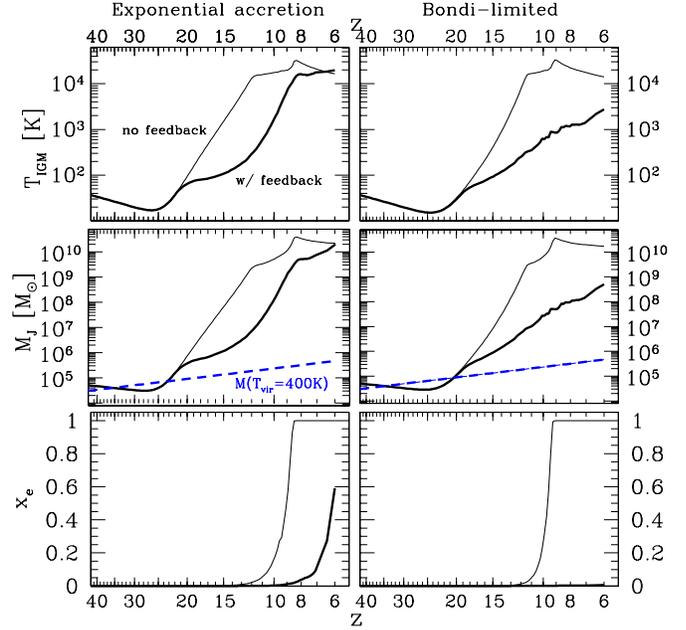}
\vspace{-0.25in}
\caption{
{\it Top panels:} The IGM temperature as a function of redshift.
{\it Middle:} The cosmological Jeans mass (black solid curves) and 
our fiducial halo mass threshold for seed BH formation (dashed blue curves).
{\it Bottom:}
The IGM electron fraction.
Thin curves represent models without feedback, whose BH mass evolution
is given in Fig. \ref{fig:nofb}; thick curves represent the models
with feedback shown in Fig. \ref{fig:withfb}. As in
Figs. \ref{fig:nofb}$-$\ref{fig:bhmf}, the exponential accretion
models are shown on the left and the Bondi-limited models on the
right.  }
\label{fig:TIGM}
\end{figure}

We show the thermal and ionization histories of the IGM in Figure
\ref{fig:TIGM}.  As with the previous figures, the exponential
accretion models are plotted on the left panels, while the
Bondi-limited models are shown on the right.  The models with and
without feedback are plotted with thick and thin curves, respectively.
Note that in the models without feedback, we still allow miniquasars
to heat and ionize the IGM, but BH formation and accretion are assumed
to be unaffected by this cosmic climate change. In particular, seed BHs
continue to form at $M_{\rm halo}^{({\rm seed})}$ ($T_{\rm
  vir}=400\K$), and BH accretion rates proceed universally at $f_{\rm
  Edd}$ in the exponential model, while proceeding as if the IGM were
cooling adiabatically in the Bondi model.  In the models with
feedback, seed formation and accretion are suppressed by the warming
of the IGM, as described above in \S\ref{subsec:withfb}.  The top pair
of panels in Figure \ref{fig:TIGM} show the IGM temperature $T_{\rm
  IGM}(z)$ for each model.  The middle pair show the corresponding
cosmological Jeans mass; the seeding mass threshold $M_{\rm
  halo}^{({\rm seed})}$ due to ${\rm H}_{2}$ cooling is plotted in
blue dashed lines alongside the Jeans mass.  The bottom panels show
the electron fraction $x_{e}=(n_{\rm HII}+n_{\rm HeII}+2n_{\rm
  HeIII})/(n_{\rm H}+2n_{\rm He})$.

Together, Figures \ref{fig:withfb} and \ref{fig:TIGM} shed light on
the coupling of the BH accretion history with the IGM temperature
history.  Initially, the evolution of $T_{\rm IGM}(z)$ and $M_{J}(z)$
follow a power-law decay due to adiabatic cooling. The buildup of the
X-ray background heats the IGM, causing a reduction in the seed BH
formation rate once $M_{J}$ becomes larger than $M_{\rm halo}^{({\rm
seed})}$.  The suppression of seed formation occurs at a unique
characteristic temperature, which may be estimated analytically as
follows.  The Jeans mass scales with the IGM temperature and density
in the same way that the halo mass scales with the virial temperature
and characteristic internal gas density (both require the
sound--crossing time to equal the dynamical time): $M\propto
(T^3/\rho)^{1/2}$.  Accounting for the gas overdensity inside
collapsed halos and differences in the conventional normalization
coefficients of order unity, this leads to $M_{J}/M_{\rm halo}\approx
60 (T_{\rm IGM}/T_{\rm vir})^{3/2}$, i.e. $M_{\rm J}\approx M_{\rm
halo}^{({\rm seed})}$ when $T_{\rm IGM} \approx 26$K.  The rising
Jeans mass quenches seed BH formation at $z\approx 18$ for both the
exponential model and the Bondi-limited model.  At these redshifts,
the total nuclear BH mass density is already $\rho_{\rm BH}\approx
10^4{\rm M_\odot Mpc^{-3}}$; however, most ($>90\%$) of this density
still consists of the original seed BHs and there are no holes above
$10^{4}\Msol$.

We may also understand the energetics of the global IGM warming, as
follows.  The increase in the IGM thermal energy density, $(3/2)n_{\rm
b} k \Delta T_{\rm IGM}$---where $n_{\rm b}$ is the baryon number
density of the IGM and $\Delta T_{\rm IGM}$ is the difference between
the actual IGM temperature and the value expected from adiabatic
cooling (eq. \ref{eq:TIGMad})---should scale with the total energy
density emitted by accreting BHs.  The latter is equal to
$\epsilon\Delta \rho_{\rm acc}c^{2}$, where $\Delta \rho_{\rm acc}$ is
the global mass density accreted by BHs.  Taking into account that
only a fraction $f_{\rm heat}$ of the accretion power goes toward
heating the IGM, we obtain
\beq \Delta T_{\rm IGM}\approx 10^{2}f_{\rm
heat}\frac{\Delta \rho_{\rm acc}}{\Msol \Mpc^{-3}}
\left(\frac{\epsilon}{0.07}\right)\left(\frac{\mu}{1.2}\right) \K.
\label{eq:fheat}
\eeq
We find that initially, at $z>20$, only $f_{\rm heat}\sim
10^{-4}-10^{-3}$ of the (mini-) quasar energy goes into heating the
IGM, but that this increases to $10^{-2}$ at $z\ltsim 8$.  The
redshift dependence of the heating efficiency is due to the fact that
not all of the hard photons are immediately absorbed by the IGM due to
their long mean free paths, which results in a delay between emission
and absorption.  Further, only $1-10\%$ of the spectrum is emitted at
$E\sim 1 \;{\rm keV}$, with more massive BHs emitting softer spectra.
Some of the energy is also lost to ionization and to the CMB through
Compton cooling.  

From the accretion history of $\rho_{\rm BH}$, we can also conclude
that {\em merely turning off the seed BH formation rate is
  insufficient to prevent overproduction.}  This is because X-ray
heating is initially inefficient, with only a fraction $f_{\rm heat}\sim
10^{-4}$ of the BH rest--mass energy at $z\sim 20$ going into
elevating the gas temperature.  The global BH density must therefore
grow to a value as large as $10^{4}\Msol \Mpc^{-3}$ for global X-ray
feedback to halt seed BH formation (at $T_{\rm IGM}\ga 30 \K$). At the
redshifts where this occurs ($z\sim 20$), the most massive BHs are at
most $\sim 10^{3}\Msol$.  If all BHs were to grow exponentially from
this point onward, by the time the most massive BHs reach
$10^{9}\Msol$, the entire population must grow by similar factors of
$\sim 10^{6}$. Even allowing for some of the BH growth to occur via
mergers, extreme overproduction of $\rho_{\rm BH}$ is inevitable.  To
prevent overproduction, either another mechanism must regulate (much
more severely) the formation of seed BHs, or BH accretion must be
(preferentially) extremely inefficient inside in low-mass halos at
later times.

In both models with feedback, the BH accretion rate is indeed reduced
preferentially inside low-mass halos.  In the exponential accretion
model, BH growth is reduced in host halos that do not continue merging
with other halos that exceed the Jeans mass.  (Note that this
suppression kicks in at $z\approx 25$, before seed BH formation rates
are affected.)  In the Bondi-limited model, IGM heating lowers the
central halo gas densities, and the corresponding Bondi rate, in the
lower-mass halos.

Unsurprisingly, the same growth models without feedback that
overproduce the massive BH population also appear to overpredict its
contribution to reionization.  Current observational constraints
indicate that reionization was still ongoing at $z\gta 7$
(e.g. \citealt{FanReview06}), and that the optical depth of the IGM to
Thomson scattering is $\tau_{\rm es}=0.088\pm0.014$, only about half
of which is accounted for by fully ionized gas between $0<z<6$
\citep[CMB measurements;][]{Komatsu+11}.
Our ``exponential'' model without feedback reionizes the Universe
by $z_{\rm r}=8.4$ and predicts an IGM Thomson optical depth of $\tau_{\rm es}\approx 0.09$,
while our ``Bondi'' model without feedback completes reionization by
$z_{\rm r}=9.1$ and predicts $\tau_{\rm es}\approx 0.11$.
These models reionize the Universe earlier than the empirical constraints,
{\it despite neglecting contributions to reionization by the stellar UV
radiation accompanying the mini-quasars BHs}.
The models with feedback, on the
other hand, fall well short of the empirical constraints
(neither completes reionization by $z=6$)
and allow room for stars to be the primary sources of reionization.
\cite{DHL04} showed that even scenarios where the contribution to
reionization by miniquasars is $\sim 10-20\%$ are consistent with the
observed unresolved soft X-ray background near $1-2$ keV.  As our
models with miniquasar feedback have lower contributions to
reionization (and also because our miniquasar spectra are softer than
the one used by those authors), we conclude that they are well within
the empirical constraints on the X-ray background
\citep[see, e.g.,][for a recent study]{Dijkstra+12}.

\subsubsection{BH merger rates}

\begin{figure}
\includegraphics[scale=0.44,angle=0]{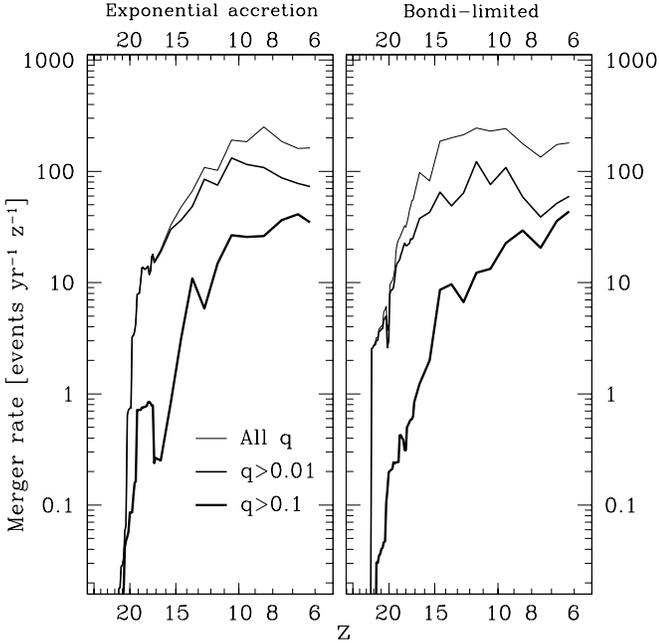}
\vspace{-0.25in}
\caption{
BH merger rates for the two assembly models with feedback
in the approximate detection window of the proposed detector
{\it eLISA}/NGO, binary mass $10^{4}\Msol<M(1+z)<10^{7}\Msol$.
The line styles demarcate the binary BH mass ratio $q\le 1$;
in order of increasing thickness:
any $q>0$, $q\ge 0.01$, and $q\ge 0.1$.
}
\label{fig:mergers}
\end{figure}

In Figure \ref{fig:mergers}, we show the number of BH mergers at $z>6$
in the binary mass window $10^{4}\Msol <M(1+z)<10^{7}\Msol$, the
estimated sensitivity window for major mergers for the proposed
gravitational-wave detector {\it eLISA}/NGO \citep{AmaroSeoane+12}.
The merger rates are shown in our two feedback models, in the previous
figures. Both models predict that $\approx 30$ major mergers (binary
mass ratios $0.1<q\le 1$) per year per unit redshift may be detectable
at $6<z<10$, with approximately twice as many events with mass ratios
$q>0.01$.  \citetalias{TH09} found that in a class of simplified
models that prevent overproduction by requiring that the $z>6$ quasar
SMBHs form from extremely rare seeds, the detection rate may be zero
over the mission lifetime.  It is encouraging that in the new models
presented here, the seeds are systematically common and the detection
rates are nonzero (see explanation for this difference below).

The majority of mergers occur with mass ratios $0.01\le q \le 0.1$
(see also \citealt{KHB+10}).  Mergers with more nearly equal masses
are relatively rare, because (i) feedback tends to increase the mass
discrepancy of BH pairs by preferentially suppressing the growth of
lower-mass BHs, (ii) BHs that survive recoil events tend to become
increasingly more massive with respect to their contemporaries, as
explained above, and (iii) halos with very different masses are
considered unable to coalesce in our merger tree, due to the long
merger timescales (see \citetalias{TH09} for details).

\subsubsection{Effects of the Lyman-Werner background}
We have also run models that included a global Lyman-Werner (LW)
background produced by the progenitor stars of the seed BHs,
which can raise the minimum halo mass threshold for subsequent seed formation.
This background and the associated mass threshold can be calculated as follows.
Approximately $10^{4}$ photons with energies $E\approx 12.9~{\rm eV}$
are produced per baryon of a progenitor star.
Through our Salpeter mass function and our progenitor-to-seed mass
relation (eq. \ref{eq:IMF}), the comoving mass density of progenitors 
is related to that of the seeds as
$\rho_{\rm PopIII}\approx 3.91\times \int_{\infty}^{0}\dot{\rho}_{\rm creat}(z)~dz$.
The fraction of the photon energy that contributes to the background,
after radiative transfer, is $\ltsim 1/10$ \citep{Haiman+97, Haiman+00}.
This gives a characteristic early LW background intensity of
\begin{align}
J_{\rm LW}\approx 6\times 10^{-23}\left(\frac{\rho_{\rm PopIII}}{10^{4}\Msol \Mpc^{-3}}\right)
\left(\frac{1+z}{21}\right)^{3}\nonumber\\
\qquad\times \erg\s^{-1}\cm^{-2}\Hz^{-1}\;{\rm sr}^{-1}.
\end{align}
Using the fitting formula from the simulations of \cite{Machacek+01},
this leads to a minimum halo mass scale for subsequent seed formation:
\beq
M_{\rm LW}\approx 2\times 10^{5}\left(\frac{\rho_{\rm PopIII}}{10^{4}\Msol \Mpc^{-3}}\right)^{1/3}
\left(\frac{1+z}{21}\right)^{-1/2}
\Msol.
\eeq
In our simulations, this mass scale is only a factor of a few above the fiducial
$T_{\rm vir}=400\K$ threshold (eq. \ref{eq:M400K})
during the brief interval $20\ltsim z\ltsim 25$, before
the X-ray global warming abruptly turns off seed formation (Fig. \ref{fig:withfb}).

The net effect of the LW background was negligible in the runs
that included both LW and X-ray feedback.
As such, and given the order-unity errors associated with LW radiative
transfer (``sawtooth'' modulation; see Haiman et al. references above)
and in the merger tree itself,
we have shown above only the results with the X-ray feedback
to best highlight the effects of the latter.

The above calculation may underestimate the early LW background,
as it does not account for the possibility of PopIII seeds forming
in groups alongside the progenitors of the BH seeds.
At later times, halos with virial temperatures above $10^{4}\K$
will efficiently form stars and add to the LW background;
however, this occurs long after the X-rays have
turned off seed formation.
Even if the LW background is much higher than what we have estimated
here, it still may not suppress BH accretion
inside previously seeded halos, due to self-shielding
(\citealt{DB96}; cf. \citealt{JWG+11} and refs. therein).

\section{Discussion}
\label{sec:discuss}

\subsection{Comparison to Other Work}
\label{subsec:compare}

X-ray heating and early reionization by miniquasars has been studied
by a number of authors.  The broader effects of various high-$z$ X-ray
backgrounds was surveyed by \citet{Oh01} and \citet{Venkat+01}, with a
special emphasis on the temperature and ionization structure of the
IGM before reionization is complete.  \cite{MBA03} used numerical
simulations to study the effects of an early soft X-ray background on
the temperature and density of gas inside minihalos at $20\la z \la
30$.  At these redshifts, the $1-2~{\rm keV}$ flux in our models is
comparable to their parameter choices $\epsilon_{X}=1$ and
$\epsilon_{X}=10$, and their results are consistent with what we find
at the onset of feedback in our models.  They report that X-ray
heating results in a moderate reduction (by a factor of a few) of the
cold gas fraction in minihalos with $M_{\rm halo}\ltsim 10^{6}\Msol$,
a mass range comparable to the heating-elevated Jeans mass in our
models.  They also found that the mean gas temperature inside less
massive halos with $M_{\rm halo}\ltsim 10^{5}\Msol$ can be elevated by
an order of magnitude, consistent with our assumption that an X-ray
background can act to suppress BH accretion in halos with $M_{\rm
halo}\ltsim M_{\rm J}$.

\citet{Madau+04} made a first attempt at specifically treating the
impact of the first BHs on the global IGM in more detail, but did not
explicitly include feedback by X-ray heating on the later generations
of BHs.  \citet{RicOst04} developed a similar semi-analytical model
for early ``pre-ionization'' by seed BHs, with the aim of identifying
a BH growth scenario which could account for the large
Thomson-scattering optical depth inferred by the earlier CMB
measurements.  Their study, while including the effect of IGM heating
on suppressing virialization below the Jeans mass, did not investigate
the corresponding modification of the resulting BH mass function.
\citet{Ricotti+05} studied X-ray pre-ionization by means of
cosmological simulations. However, their focus was again on the
observational signatures in the CMB and 21-cm signal, rather than on
the growth of SMBHs.

More recently, \citet{Devecchi+12} explored the formation of nuclear
star clusters and BH seeds, tracking the chemical, radiative and
mechanical feedback of stars on the baryonic component of the evolving
halos. In particular, they examined the role of the LW
photons in suppressing star formation, but did not include the rise of
the Jeans mass due to X-ray heating of the IGM by the seeds BHs.
Feedback effects from the LW photons in shaping BH growth in
the most massive ($10^{12-13}{\rm M_\odot}$) halos at $z=6$ have also
been recently examined in detail by \cite{Petri+12} and \cite{Agarwal+12},
using merger trees and N-body simulations, respectively;
however, those studies did not examine feedback on the lower-mass BHs in smaller halos.  In
principle, LW radiation could help alleviate the SMBH
overprediction problem, by disabling the cooling in low-mass halos at
very early times \citep{HaimanBryan06} - this is similar to the X-ray
background, although likely restricted to lower-mass minihalos 
\citep[e.g.,][]{Machacek+01}.

Finally, X-rays from miniquasars can also exert positive feedback by
triggering the formation of molecular hydrogen
\citep{Haiman+00,MBA03,KM05,Jeon+12}. For example, \citet{MBA03},
\citet{KM05} and \citet{Jeon+12} found that the X-rays can {\it
increase} the amount of cold and dense gas (and thus star formation)
in the cores of massive minihalos in their simulations.  This runs
contrary to the assumption in our ``Bondi-limited'' model that X-rays
cause a warming of the cores of low-mass halos.\footnote{If the
bottleneck on BH accretion is not the central cold gas density but
rather the external supply of gas, then the X-ray-induced suppression
of baryonic infall found by \cite{MBA03}, \cite{KM05} and others
should lead to a negative feedback on BH growth, as assumed in our
``exponential growth'' model.}  However, the results of \cite{KM05}
suggest that at sufficiently high X-ray fluxes, the heating outweighs
the enhanced ${\rm H_2}$ cooling and the feedback on the cold gas
density becomes negative.  The value of the X-ray flux where this
turnover occurs in their simulation---in minihalos in close
proximity of their miniquasar; $\epsilon_{X}\sim {\rm a~few}$ at $\sim
1~{\rm keV}$ in the normalization convention of \cite{MBA03}---is
reached in our models at $z\sim 25-20$ when $\rho_{\rm BH}\ga 10^{3}
\Msol \Mpc^{-3}$, and exceeded soon thereafter.  This suggests that
the X-ray background fluxes in our models are large enough to induce
negative feedback.  Furthermore, the $E\gta 1$ keV X-rays have mean
free paths of $\sim 1~{\rm Gpc}$.  The simulation volumes of $\sim
1~{\rm Mpc}^{3}$ employed by \cite{KM05} and \cite{Jeon+12} do not
account for the X-rays originating from sources outside the box and
thus do not address the early buildup of a global X-ray background.
(\citealt{KM05} turn on a single miniquasar at $z=21$.) Finally, even
if the X-rays exert a positive feedback by increasing the ${\rm H_2}$
cooling, the net effect may also be negative in the presence of a
strong LW field accompanying the X-rays.  Whether, and to what degree,
X-rays can induce positive feedback on structure formation and early
BH growth remains very much an open question.

\subsection{General Comments on Viable  SMBH Growth Scenarios}
\label{subsec:SMBHgrowth}

The puzzle of the origins of the $z\approx 6-7$ quasar SMBHs has
motivated numerous recent theoretical investigations.  Here, remark on
possible SMBH assembly models, attempting to generalize the results we
obtained above.

\subsubsection{Explaining the monsters}

First, what is necessary to explain the individual SMBHs?  We find
that for the most massive BHs in our simulations to grow to $M\sim
10^{9}\Msol$ by $z\approx 7$ (i.e., to explain the
\cite{Mortlock+11} quasar observation), a seed BH left behind by a
PopIII star and starting at $100\Msol$ at $z\approx 40$ must have an
effective e-folding time of $\ltsim 55 \Myr$.  That is, the mean
Eddington ratio must be $f_{\rm Edd}\ga 0.6$ for radiative efficiency
$\epsilon =0.07$. Or, more generally, these two quantities must
satisfy $f_{\rm Edd}(1-\epsilon)/\epsilon \ga 8$.
As the elapsed time
between $z\approx 40$ and $z\approx 7.08$ is $\approx 700 \Myr$, this
lower limit corresponds to accretion-driven growth by a factor of
${\rm a~few}\times 10^{5}$.  Our models show that mergers can
contribute to assembly by an additional factor $10-100$.
Note that 
for the progenitors of the $z\sim 6$ quasar SMBHs, this requires
that the mean accretion be close to ,
that the radiative efficiency be no larger than $\sim 0.1$, or both.
Note that the above requirement suggests that the progenitors
of the $z>6$ quasar SMBHs have (averaged over their growth histories)
accretion rates close to or exceeding the Eddington limit,
relatively low radiative efficiencies $\epsilon\ltsim 0.1$, or both.
Such low radiative efficiencies suggest that the BHs cannot
be spinning rapidly, consistent with accretion disk models with
magnetohydrodynamic turbulence
\citep[see, e.g.][and references therein]{Shapiro05}.
A hypothetical, single BH accreting at the Eddington limit and
$\epsilon=0.07$ can grow by a factor $\approx 10^{9}$ in the same
redshift interval.

Another possibility is that the ancestors of the monster SMBHs were
born with much higher masses.  For example, particularly massive BHs
with $M_{\rm BH}>10^{4}\Msol$ could have formed in halos whose gaseous
components have unusually low angular momentum
\citep{Koushiappas+04,LodatoNatarajan06}; in halos that are heavily
irradiated by massive neighbors \citep{Dijkstra+08}; in environments
with high ambient magnetic fields \citep{Sethi+10}; or in rare
instances of ionizing shocks in dense halo cores \citep{IO12}.

A common minimum requirement for both classes of seed models is that a
$\sim 10^{5}\Msol$ BH be in place by $z\approx10$; such BHs can grow
at the Eddington rate to $\sim 10^{9}\Msol$ by $z\approx 7$.  The
recent work by \cite{DiMatteo+12} suggests that once such a BH is in
place, then cold gas accretion by the rapidly growing host halo could
help deliver the gas, at least on large scales, to sustain its
near-Eddington growth.  Any seed model that satisfies this condition
can, in principle, explain the individual quasar SMBH masses observed
to date.  The above requirement also suggest a fundamental degeneracy
for assembly models of high-redshift SMBHs in the $6\ltsim z\ltsim 10$
range.

\subsubsection{Modeling the overall population}

In light of the above points, the greater challenge arguably lies in
distinguishing between the various assembly scenarios.  Probing the
non-degenerate parameter space $(M_{\rm BH}<10^{5}\Msol, z\ga 10)$
through direct electromagnetic observation will be difficult. This is
about the detection sensitivity of the {\it James Webb Space
  Telescope}, assuming Eddington luminosity and a $\sim 10^{5}\s$
integration \citep{Haiman12}.  If IGM preheating suppresses star
formation in halos with $M_{\rm halo}\ltsim 10^{9}\Msol$, this may
lead to an observable faint-end cutoff in the high-redshift galaxy
luminosity function \citep{BarkanaLoeb00} and a drop in the rate of
high-redshift supernovae \citep{Mesinger+06a}; both effects could be
observed by the {\it JWST} and provide circumstantial evidence for
negative feedback on low-mass halos. However, the most obvious
candidate for distinguishing models is a gravitational-wave detector
such as {\it eLISA}/NGO, which could directly constrain BH merger
rates, masses and abundances at high-redshift.

Previous studies had suggested that to avoid overproduction, seed BHs
must be rare (\citealt{Bromley+04}; \citetalias{TH09}).  In principle,
the occupation fraction of SMBHs in galaxies can approach unity by low
redshift even if nuclear BHs were very rare at early times
\citep{Menou+01,Lippai+09}.  If seeds are indeed rare, then it is
conceivable that the BH merger rate at high redshift $z\ga 6$) would
be so low that no detectable events will be expected by {\it
  eLISA}/NGO (\citetalias{TH09}).  A rare seed BH population would
also have a smaller impact via global feedback.  Therefore, one way to
indirectly confirm such models is by ruling out alternative scenarios,
with common seed BHs, through null detections.

The alternative possibility proposed here is that seeds are
systematically much more common, but that BHs grow much less rapidly
in low-mass halos at late times. This could be because AGN feedback or
other local processes self-regulate the BH mass with respect to the
host galaxy properties, i.e. processes similar to those resulting in
the locally observed $M-\sigma$ relation (\citetalias{TH09},
\citealt{VN09}).  In this paper, we have shown that similar
effects can be produced instead by global feedback; the hard photons
from the early accreting BHs can heat the IGM and regulate subsequent
BH formation and growth.  The same BHs that go on to assemble the
$z\sim 6$ SMBHs can slow the growth of the low end of the BH mass
function and avoid overproduction.  One important consequence of the
common-seed scenarios, regardless of whether the mechanism regulating
BH growth is local or global, is that the mergers of the SMBH
ancestors and their intermediate-mass contemporaries should be
frequent enough to be detectable by an observatory such as {\it
  eLISA}/NGO (cf. \citealt{Sesana+07}, \citealt{Micic+07}).

Additionally, if the seeds are common, then the X-rays emitted in the
course of their mass growth could generically influence cosmological
structures.  This may be inevitable, regardless of whether the first
nuclear BHs had masses of $10\Msol$ or $10^{5}\Msol$; en route to
becoming $10^{7-9}\Msol$ SMBHs, the vast majority of the mass growth
must occur through accretion.
Indeed, \ref{eq:fheat} suggests that the IGM will be heated
to $\sim 10^{4}\K$ if the comoving SMBH mass density at $z=6$
is several percent of the present-day value,
and if $\sim 0.1\%$ of the miniquasar energy output required
to build up this mass density goes toward heating the IGM.

Our models predict that miniquasars
play a significant role in establishing the cosmological temperature
history of the IGM, since X-rays from miniquasars would heat the early
Universe much more effectively than UV radiation from PopIII stars
\citep{Madau+04, Ciardi+10}.  Our results suggest that PopIII star formation may
be affected by miniquasars as early as $z\approx 20$.  Other sources,
such as LW radiation from PopIII stars and the accreting BHs,
hard photons from non-nuclear stellar-mass BHs
\citep{WheelerJohnson11}
and high-mass X-ray binaries \citep{Mirabel+11},
or local AGN feedback and regulation,
may also have contributed to this feedback.

\section{Conclusions}
\label{sec:conclude}

Observations of high-redshift quasars at $z\gta 6$ imply that SMBHs
with masses $M\gta10^{9}\Msol$ were in place as early as $z>7$.  In
this paper, we considered models in which these SMBHs grow from
stellar-mass seed BHs forming at ultra-high redshifts (i.e. the
remnants of first-generation stars at $z\sim 30-40$), obeying the
Eddington limit on the mass accretion rate.  Previous work has shown
that this is feasible, but only if the duty cycle for accretion is of
order unity.

Here we highlighted a relatively under-appreciated problem in this
class of models: unless the growth of BHs in low-mass halos is
preferentially and severely suppressed, the models overpredict the
abundance of $10^{5-7} {\rm M_\odot}$ BHs in galactic nuclei by
several orders of magnitude.

Using Monte-Carlo realizations of the merger and growth history of
BHs, we show that the X-rays emitted by the earliest accreting BHs can
heat the IGM, and suppress the formation and growth of subsequent
generations of BHs in low-mass halos.  In this ``global warming''
scenario, the BHs originally responsible for the warming are largely
unaffected by it, because they reside in the most massive halos, well
above the Jeans mass, and frequently merge with other massive halos
with cold gas, facilitating BH growth.  However, the negative effects
are felt by the next generations of low-mass halos, in which seed
formation and BH accretion are suppressed.

We presented specific models with global miniquasar feedback that
provide excellent agreement with recent estimates of the $z=6$ SMBH
mass function.  These models could be constrained through direct
observations by {\em JWST}, and through the detection of tens of BH
mergers at $z>6$ by the proposed gravitational-wave observatory {\it
  eLISA}/NGO.

We compared our work to previous studies investigating the effects of
X-ray feedback on early structure formation.  A primary uncertainty is
whether, and to what degree, a moderate X-ray background may induce
{\it positive} feedback on the cold gas content of minihalos through
enhanced ${\rm H_2}$ formation and cooling, and whether this effect
outweighs the heating of the gas by the X-rays.  Finally, a limitation
of our merger tree approach is that we do not consider local effects
due to clustering and to proximity to miniquasars, which may be
significant \citep[e.g.,][]{KM05}.  While our work serves as a
proof-of-concept that global X-ray feedback from accreting BHs can
regulate their growth, more detailed studies on this subject are
warranted.

\section*{ACKNOWLEDGMENTS}

We thank Jerry Ostriker and Greg Bryan for insightful conversations,
and Daniel Mortlock for useful discussions about the $z=7.08$ quasar.
RP and ZH thank the Centro de Ciencias de Benasque Pedro Pascual,
where this project was initiated.  This work was partially supported
by NSF grant AST-1009396 (to RP) and by NASA grant NNX11AE05G (to ZH).

\bibliographystyle{mn2e}

\begin{thebibliography}{140}
\expandafter\ifx\csname natexlab\endcsname\relax\def\natexlab#1{#1}\fi

\bibitem[{{Abel}, {Bryan} \& {Norman}(2002){Abel}, {Bryan}, \&
  {Norman}}]{Abel+02}
{Abel} T., {Bryan} G.~L., {Norman} M.~L., 2002, Science, 295, 93

\bibitem[{{Agarwal} {et~al}\mbox{.}(2012){Agarwal}, {Khochfar}, {Johnson},
  {Neistein}, {Dalla Vecchia}, \& {Livio}}]{Agarwal+12}
{Agarwal} B., {Khochfar} S., {Johnson} J.~L., {Neistein} E., {Dalla Vecchia}
  C., {Livio} M., 2012, arXiv e-prints 1205.6464

\bibitem[{{Ahn} \& {Shapiro}(2007)}]{AhnShapiro07}
{Ahn} K., {Shapiro} P.~R., 2007, \mnras, 375, 881

\bibitem[{{Alvarez}, {Wise} \& {Abel}(2009){Alvarez}, {Wise}, \&
  {Abel}}]{Alvarez+09}
{Alvarez} M.~A., {Wise} J.~H., {Abel} T., 2009, \apjl, 701, L133

\bibitem[{{Amaro-Seoane} {et~al}\mbox{.}(2012){Amaro-Seoane}, {Aoudia},
  {Babak}, {Bin{\'e}truy}, {Berti}, {Boh{\'e}}, {Caprini}, {Colpi}, {Cornish},
  {Danzmann}, {Dufaux}, {Gair}, {Jennrich}, {Jetzer}, {Klein}, {Lang}, {Lobo},
  {Littenberg}, {McWilliams}, {Nelemans}, {Petiteau}, {Porter}, {Schutz},
  {Sesana}, {Stebbins}, {Sumner}, {Vallisneri}, {Vitale}, {Volonteri}, \&
  {Ward}}]{AmaroSeoane+12}
{Amaro-Seoane} P. {et~al.}, 2012, arXiv e-prints 1201.3621

\bibitem[{{Angulo} {et~al}\mbox{.}(2012){Angulo}, {Springel}, {White}, {Cole},
  {Jenkins}, {Baugh}, \& {Frenk}}]{Angulo+12}
{Angulo} R.~E., {Springel} V., {White} S.~D.~M., {Cole} S., {Jenkins} A.,
  {Baugh} C.~M., {Frenk} C.~S., 2012, arXiv e-prints 1203.5339

\bibitem[{{Barkana} \& {Loeb}(2000)}]{BarkanaLoeb00}
{Barkana} R., {Loeb} A., 2000, \apj, 539, 20

\bibitem[{{Barkana} \& {Loeb}(2001)}]{BarkanaLoeb01}
---, 2001, \physrep, 349, 125

\bibitem[{{Begelman}, {Volonteri} \& {Rees}(2006){Begelman}, {Volonteri}, \&
  {Rees}}]{Begelman+06}
{Begelman} M.~C., {Volonteri} M., {Rees} M.~J., 2006, \mnras, 370, 289

\bibitem[{{Blaes}(2004)}]{Blaes04}
{Blaes} O.~M., 2004, in Accretion Discs, Jets and High Energy Phenomena in
  Astrophysics, {V.~Beskin, G.~Henri, F.~Menard, G.~Pelletier, J.~Dalibard },
  ed., Springer Publishing Company, New York, NY, USA, pp. 137--185

\bibitem[{{Blecha} \& {Loeb}(2008)}]{BlechaLoeb08}
{Blecha} L., {Loeb} A., 2008, \mnras, 390, 1311

\bibitem[{{Bogdanovi{\'c}}, {Reynolds} \& {Miller}(2007){Bogdanovi{\'c}},
  {Reynolds}, \& {Miller}}]{Bogdanovic+07}
{Bogdanovi{\'c}} T., {Reynolds} C.~S., {Miller} M.~C., 2007, \apjl, 661, L147

\bibitem[{{Bond}, {Arnett} \& {Carr}(1984){Bond}, {Arnett}, \&
  {Carr}}]{Bond+84}
{Bond} J.~R., {Arnett} W.~D., {Carr} B.~J., 1984, \apj, 280, 825

\bibitem[{{Bromley}, {Somerville} \& {Fabian}(2004){Bromley}, {Somerville}, \&
  {Fabian}}]{Bromley+04}
{Bromley} J.~M., {Somerville} R.~S., {Fabian} A.~C., 2004, \mnras, 350, 456

\bibitem[{{Bromm}, {Coppi} \& {Larson}(2002){Bromm}, {Coppi}, \&
  {Larson}}]{Bromm+02}
{Bromm} V., {Coppi} P.~S., {Larson} R.~B., 2002, \apj, 564, 23

\bibitem[{{Bromm} \& {Loeb}(2003)}]{BrommLoeb03}
{Bromm} V., {Loeb} A., 2003, \apj, 596, 34

\bibitem[{{Bullock} {et~al}\mbox{.}(2001){Bullock}, {Kolatt}, {Sigad},
  {Somerville}, {Kravtsov}, {Klypin}, {Primack}, \& {Dekel}}]{Bullock+01}
{Bullock} J.~S., {Kolatt} T.~S., {Sigad} Y., {Somerville} R.~S., {Kravtsov}
  A.~V., {Klypin} A.~A., {Primack} J.~R., {Dekel} A., 2001, \mnras, 321, 559

\bibitem[{{Ciardi}, {Salvaterra} \& {Di Matteo}(2010){Ciardi}, {Salvaterra}, \&
  {Di Matteo}}]{Ciardi+10}
{Ciardi} B., {Salvaterra} R., {Di Matteo} T., 2010, \mnras, 401, 2635

\bibitem[{{Ciotti} \& {Ostriker}(2001)}]{CO01}
{Ciotti} L., {Ostriker} J.~P., 2001, \apj, 551, 131

\bibitem[{{Cuadra} {et~al}\mbox{.}(2009){Cuadra}, {Armitage}, {Alexander}, \&
  {Begelman}}]{Cuadra+09}
{Cuadra} J., {Armitage} P.~J., {Alexander} R.~D., {Begelman} M.~C., 2009,
  \mnras, 393, 1423

\bibitem[{{Devecchi} \& {Volonteri}(2009)}]{DV09}
{Devecchi} B., {Volonteri} M., 2009, \apj, 694, 302

\bibitem[{{Devecchi} {et~al}\mbox{.}(2012){Devecchi}, {Volonteri}, {Rossi},
  {Colpi}, \& {Portegies Zwart}}]{Devecchi+12}
{Devecchi} B., {Volonteri} M., {Rossi} E.~M., {Colpi} M., {Portegies Zwart} S.,
  2012, \mnras, 421, 1465

\bibitem[{{Di Matteo} {et~al}\mbox{.}(2012){Di Matteo}, {Khandai}, {DeGraf},
  {Feng}, {Croft}, {Lopez}, \& {Springel}}]{DiMatteo+12}
{Di Matteo} T., {Khandai} N., {DeGraf} C., {Feng} Y., {Croft} R.~A.~C., {Lopez}
  J., {Springel} V., 2012, \apjl, 745, L29

\bibitem[{{Dijkstra} {et~al}\mbox{.}(2012){Dijkstra}, {Gilfanov}, {Loeb}, \&
  {Sunyaev}}]{Dijkstra+12}
{Dijkstra} M., {Gilfanov} M., {Loeb} A., {Sunyaev} R., 2012, \mnras, 421, 213

\bibitem[{{Dijkstra}, {Haiman} \& {Loeb}(2004){Dijkstra}, {Haiman}, \&
  {Loeb}}]{DHL04}
{Dijkstra} M., {Haiman} Z., {Loeb} A., 2004, \apj, 613, 646

\bibitem[{{Dijkstra} {et~al}\mbox{.}(2008){Dijkstra}, {Haiman}, {Mesinger}, \&
  {Wyithe}}]{Dijkstra+08}
{Dijkstra} M., {Haiman} Z., {Mesinger} A., {Wyithe} J.~S.~B., 2008, \mnras,
  391, 1961

\bibitem[{{Dotti} {et~al}\mbox{.}(2009){Dotti}, {Montuori}, {Decarli},
  {Volonteri}, {Colpi}, \& {Haardt}}]{Dotti+09}
{Dotti} M., {Montuori} C., {Decarli} R., {Volonteri} M., {Colpi} M., {Haardt}
  F., 2009, \mnras, 398, L73

\bibitem[{{Draine} \& {Bertoldi}(1996)}]{DB96}
{Draine} B.~T., {Bertoldi} F., 1996, \apj, 468, 269

\bibitem[{{Escala} {et~al}\mbox{.}(2004){Escala}, {Larson}, {Coppi}, \&
  {Mardones}}]{Escala+04}
{Escala} A., {Larson} R.~B., {Coppi} P.~S., {Mardones} D., 2004, \apj, 607, 765

\bibitem[{{Fan}(2006)}]{Fan06}
{Fan} X., 2006, \nar, 50, 665

\bibitem[{{Fan}, {Carilli} \& {Keating}(2006){Fan}, {Carilli}, \&
  {Keating}}]{FanReview06}
{Fan} X., {Carilli} C.~L., {Keating} B., 2006, \araa, 44, 415

\bibitem[{{Furlanetto} \& {Stoever}(2010)}]{FS10}
{Furlanetto} S.~R., {Stoever} S.~J., 2010, \mnras, 404, 1869

\bibitem[{{Gnedin}(2000)}]{Gnedin00}
{Gnedin} N.~Y., 2000, \apj, 542, 535

\bibitem[{{Greif} {et~al}\mbox{.}(2012){Greif}, {Bromm}, {Clark}, {Glover},
  {Smith}, {Klessen}, {Yoshida}, \& {Springel}}]{Greif+12}
{Greif} T.~H., {Bromm} V., {Clark} P.~C., {Glover} S.~C.~O., {Smith} R.~J.,
  {Klessen} R.~S., {Yoshida} N., {Springel} V., 2012, \mnras, 424, 399

\bibitem[{{Greif} {et~al}\mbox{.}(2011{\natexlab{a}}){Greif}, {White},
  {Klessen}, \& {Springel}}]{Greif+11a}
{Greif} T.~H., {White} S.~D.~M., {Klessen} R.~S., {Springel} V.,
  2011{\natexlab{a}}, \apj, 736, 147

\bibitem[{{Greif} {et~al}\mbox{.}(2011{\natexlab{b}}){Greif}, {Springel},
  {White}, {Glover}, {Clark}, {Smith}, {Klessen}, \& {Bromm}}]{Greif+11b}
{Greif} T.~H., {Springel} V., {White} S.~D.~M., {Glover} S.~C.~O., {Clark}
  P.~C., {Smith} R.~J., {Klessen} R.~S., {Bromm} V., 2011{\natexlab{b}}, \apj,
  737, 75

\bibitem[{{Guedes} {et~al}\mbox{.}(2009){Guedes}, {Madau}, {Kuhlen}, {Diemand},
  \& {Zemp}}]{Guedes+09}
{Guedes} J., {Madau} P., {Kuhlen} M., {Diemand} J., {Zemp} M., 2009, \apj, 702,
  890

\bibitem[{{Haardt} \& {Madau}(1996)}]{HaardtMadau96}
{Haardt} F., {Madau} P., 1996, \apj, 461, 20

\bibitem[{{Haehnelt}(2003)}]{Haehnelt03}
{Haehnelt} M.~G., 2003, \CQG, 20, 31

\bibitem[{{Haehnelt} \& {Rees}(1993)}]{HR93}
{Haehnelt} M.~G., {Rees} M.~J., 1993, \mnras, 263, 168

\bibitem[{{Haiman}(2004)}]{Haiman04}
{Haiman} Z., 2004, \apj, 613, 36

\bibitem[{{Haiman}(2012)}]{Haiman12}
---, 2012, arXiv e-prints 1203.6075

\bibitem[{{Haiman}, {Abel} \& {Rees}(2000){Haiman}, {Abel}, \&
  {Rees}}]{Haiman+00}
{Haiman} Z., {Abel} T., {Rees} M.~J., 2000, \apj, 534, 11

\bibitem[{{Haiman} \& {Bryan}(2006)}]{HaimanBryan06}
{Haiman} Z., {Bryan} G.~L., 2006, \apj, 650, 7

\bibitem[{{Haiman} \& {Loeb}(2001)}]{HaimanLoeb01}
{Haiman} Z., {Loeb} A., 2001, \apj, 552, 459

\bibitem[{{Haiman}, {Rees} \& {Loeb}(1997){Haiman}, {Rees}, \&
  {Loeb}}]{Haiman+97}
{Haiman} Z., {Rees} M.~J., {Loeb} A., 1997, \apj, 476, 458

\bibitem[{{Haiman}, {Thoul} \& {Loeb}(1996){Haiman}, {Thoul}, \&
  {Loeb}}]{Haiman+96}
{Haiman} Z., {Thoul} A.~A., {Loeb} A., 1996, \apj, 464, 523

\bibitem[{{Heger} {et~al}\mbox{.}(2003){Heger}, {Fryer}, {Woosley}, {Langer},
  \& {Hartmann}}]{Heger+03}
{Heger} A., {Fryer} C.~L., {Woosley} S.~E., {Langer} N., {Hartmann} D.~H.,
  2003, \apj, 591, 288

\bibitem[{{Holley-Bockelmann} {et~al}\mbox{.}(2010){Holley-Bockelmann},
  {Micic}, {Sigurdsson}, \& {Rubbo}}]{KHB+10}
{Holley-Bockelmann} K., {Micic} M., {Sigurdsson} S., {Rubbo} L.~J., 2010, \apj,
  713, 1016

\bibitem[{{Hopkins}, {Richards} \& {Hernquist}(2007){Hopkins}, {Richards}, \&
  {Hernquist}}]{Hopkins+07b}
{Hopkins} P.~F., {Richards} G.~T., {Hernquist} L., 2007, \apj, 654, 731

\bibitem[{{Hosokawa}, {Omukai} \& {Yorke}(2012){Hosokawa}, {Omukai}, \&
  {Yorke}}]{Hosokawa+12}
{Hosokawa} T., {Omukai} K., {Yorke} H.~W., 2012, arXiv e-prints 1203.2613

\bibitem[{{Hosokawa} {et~al}\mbox{.}(2011){Hosokawa}, {Omukai}, {Yoshida}, \&
  {Yorke}}]{Hosokawa+11}
{Hosokawa} T., {Omukai} K., {Yoshida} N., {Yorke} H.~W., 2011, Science, 334,
  1250

\bibitem[{{Inayoshi} \& {Omukai}(2012)}]{IO12}
{Inayoshi} K., {Omukai} K., 2012, \mnras, 422, 2539

\bibitem[{{Jarosik} {et~al}\mbox{.}(2011){Jarosik}, {Bennett}, {Dunkley},
  {Gold}, {Greason}, {Halpern}, {Hill}, {Hinshaw}, {Kogut}, {Komatsu},
  {Larson}, {Limon}, {Meyer}, {Nolta}, {Odegard}, {Page}, {Smith}, {Spergel},
  {Tucker}, {Weiland}, {Wollack}, \& {Wright}}]{Jarosik+11}
{Jarosik} N. {et~al.}, 2011, \apjs, 192, 14

\bibitem[{{Jeon} {et~al}\mbox{.}(2012){Jeon}, {Pawlik}, {Greif}, {Glover},
  {Bromm}, {Milosavljevi{\'c}}, \& {Klessen}}]{Jeon+12}
{Jeon} M., {Pawlik} A.~H., {Greif} T.~H., {Glover} S.~C.~O., {Bromm} V.,
  {Milosavljevi{\'c}} M., {Klessen} R.~S., 2012, \apj, 754, 34

\bibitem[{{Kelly} {et~al}\mbox{.}(2010){Kelly}, {Vestergaard}, {Fan},
  {Hopkins}, {Hernquist}, \& {Siemiginowska}}]{Kelly+10}
{Kelly} B.~C., {Vestergaard} M., {Fan} X., {Hopkins} P., {Hernquist} L.,
  {Siemiginowska} A., 2010, \apj, 719, 1315

\bibitem[{{Klypin}, {Trujillo-Gomez} \& {Primack}(2011){Klypin},
  {Trujillo-Gomez}, \& {Primack}}]{Klypin+11}
{Klypin} A.~A., {Trujillo-Gomez} S., {Primack} J., 2011, \apj, 740, 102

\bibitem[{{Kollmeier} {et~al}\mbox{.}(2006){Kollmeier}, {Onken}, {Kochanek},
  {Gould}, {Weinberg}, {Dietrich}, {Cool}, {Dey}, {Eisenstein}, {Jannuzi}, {Le
  Floc'h}, \& {Stern}}]{Kollmeier+06}
{Kollmeier} J.~A. {et~al.}, 2006, \apj, 648, 128

\bibitem[{{Komatsu} {et~al}\mbox{.}(2011){Komatsu}, {Smith}, {Dunkley},
  {Bennett}, {Gold}, {Hinshaw}, {Jarosik}, {Larson}, {Nolta}, {Page},
  {Spergel}, {Halpern}, {Hill}, {Kogut}, {Limon}, {Meyer}, {Odegard}, {Tucker},
  {Weiland}, {Wollack}, \& {Wright}}]{Komatsu+11}
{Komatsu} E. {et~al.}, 2011, \apjs, 192, 18

\bibitem[{{Koushiappas}, {Bullock} \& {Dekel}(2004){Koushiappas}, {Bullock}, \&
  {Dekel}}]{Koushiappas+04}
{Koushiappas} S.~M., {Bullock} J.~S., {Dekel} A., 2004, \mnras, 354, 292

\bibitem[{{Kramer}, {Haiman} \& {Oh}(2006){Kramer}, {Haiman}, \&
  {Oh}}]{Kramer+06}
{Kramer} R.~H., {Haiman} Z., {Oh} S.~P., 2006, \apj, 649, 570

\bibitem[{{Kuhlen} \& {Madau}(2005)}]{KM05}
{Kuhlen} M., {Madau} P., 2005, \mnras, 363, 1069

\bibitem[{{Kulkarni} \& {Loeb}(2012)}]{KulkarniLoeb12}
{Kulkarni} G., {Loeb} A., 2012, \mnras, 422, 1306

\bibitem[{{Lacey} \& {Cole}(1993)}]{LaceyCole93}
{Lacey} C., {Cole} S., 1993, \mnras, 262, 627

\bibitem[{{Lawrence}(1991)}]{Lawrence91}
{Lawrence} A., 1991, \mnras, 252, 586

\bibitem[{{Lippai}, {Frei} \& {Haiman}(2009){Lippai}, {Frei}, \&
  {Haiman}}]{Lippai+09}
{Lippai} Z., {Frei} Z., {Haiman} Z., 2009, \apj, 701, 360

\bibitem[{{Lodato} \& {Natarajan}(2006)}]{LodatoNatarajan06}
{Lodato} G., {Natarajan} P., 2006, \mnras, 371, 1813

\bibitem[{{Loeb} \& {Rasio}(1994)}]{LoebRasio94}
{Loeb} A., {Rasio} F.~A., 1994, \apj, 432, 52

\bibitem[{{Lousto} {et~al}\mbox{.}(2010){Lousto}, {Campanelli}, {Zlochower}, \&
  {Nakano}}]{Lousto+10}
{Lousto} C.~O., {Campanelli} M., {Zlochower} Y., {Nakano} H., 2010, \CQG, 27,
  114006

\bibitem[{{Machacek}, {Bryan} \& {Abel}(2001){Machacek}, {Bryan}, \&
  {Abel}}]{Machacek+01}
{Machacek} M.~E., {Bryan} G.~L., {Abel} T., 2001, \apj, 548, 509

\bibitem[{{Machacek}, {Bryan} \& {Abel}(2003){Machacek}, {Bryan}, \&
  {Abel}}]{MBA03}
---, 2003, \mnras, 338, 273

\bibitem[{{Madau}, {Ferrara} \& {Rees}(2001){Madau}, {Ferrara}, \&
  {Rees}}]{Madau+01}
{Madau} P., {Ferrara} A., {Rees} M.~J., 2001, \apj, 555, 92

\bibitem[{{Madau} {et~al}\mbox{.}(2004){Madau}, {Rees}, {Volonteri}, {Haardt},
  \& {Oh}}]{Madau+04}
{Madau} P., {Rees} M.~J., {Volonteri} M., {Haardt} F., {Oh} S.~P., 2004, \apj,
  604, 484

\bibitem[{{Marconi} {et~al}\mbox{.}(2004){Marconi}, {Risaliti}, {Gilli},
  {Hunt}, {Maiolino}, \& {Salvati}}]{Marconi+04}
{Marconi} A., {Risaliti} G., {Gilli} R., {Hunt} L.~K., {Maiolino} R., {Salvati}
  M., 2004, \mnras, 351, 169

\bibitem[{{Martini}(2004)}]{Martini04}
{Martini} P., 2004, in Coevolution of Black Holes and Galaxies, {L.~C.~Ho},
  ed., {Cambridge University Press, Cambridge, United Kingdom}, pp. 169--+

\bibitem[{{Menou}, {Haiman} \& {Narayanan}(2001){Menou}, {Haiman}, \&
  {Narayanan}}]{Menou+01}
{Menou} K., {Haiman} Z., {Narayanan} V.~K., 2001, \apj, 558, 535

\bibitem[{{Merloni} \& {Heinz}(2008)}]{MerloniHeinz08}
{Merloni} A., {Heinz} S., 2008, \mnras, 388, 1011

\bibitem[{{Mesinger}, {Johnson} \& {Haiman}(2006){Mesinger}, {Johnson}, \&
  {Haiman}}]{Mesinger+06a}
{Mesinger} A., {Johnson} B.~D., {Haiman} Z., 2006, \apj, 637, 80

\bibitem[{{Micic} {et~al}\mbox{.}(2007){Micic}, {Holley-Bockelmann},
  {Sigurdsson}, \& {Abel}}]{Micic+07}
{Micic} M., {Holley-Bockelmann} K., {Sigurdsson} S., {Abel} T., 2007, \mnras,
  380, 1533

\bibitem[{{Milosavljevi{\'c}} {et~al}\mbox{.}(2009){Milosavljevi{\'c}},
  {Bromm}, {Couch}, \& {Oh}}]{Milos+09}
{Milosavljevi{\'c}} M., {Bromm} V., {Couch} S.~M., {Oh} S.~P., 2009, \apj, 698,
  766

\bibitem[{{Mirabel} {et~al}\mbox{.}(2011){Mirabel}, {Dijkstra}, {Laurent},
  {Loeb}, \& {Pritchard}}]{Mirabel+11}
{Mirabel} I.~F., {Dijkstra} M., {Laurent} P., {Loeb} A., {Pritchard} J.~R.,
  2011, \aap, 528, A149

\bibitem[{{Mortlock} {et~al}\mbox{.}(2011){Mortlock}, {Warren}, {Venemans},
  {Patel}, {Hewett}, {McMahon}, {Simpson}, {Theuns}, {Gonz{\'a}les-Solares},
  {Adamson}, {Dye}, {Hambly}, {Hirst}, {Irwin}, {Kuiper}, {Lawrence}, \&
  {R{\"o}ttgering}}]{Mortlock+11}
{Mortlock} D.~J. {et~al.}, 2011, \nat, 474, 616

\bibitem[{{Naoz} \& {Barkana}(2007)}]{NaozBark07}
{Naoz} S., {Barkana} R., 2007, \mnras, 377, 667

\bibitem[{{Navarro}, {Frenk} \& {White}(1996){Navarro}, {Frenk}, \&
  {White}}]{NFW96}
{Navarro} J.~F., {Frenk} C.~S., {White} S.~D.~M., 1996, \apj, 462, 563

\bibitem[{{Oh}(2001)}]{Oh01}
{Oh} S.~P., 2001, \apj, 553, 499

\bibitem[{{Oh} \& {Haiman}(2002)}]{OH02}
{Oh} S.~P., {Haiman} Z., 2002, \apj, 569, 558

\bibitem[{{Oh} \& {Haiman}(2003)}]{OH03}
---, 2003, \mnras, 346, 456

\bibitem[{{Ohkubo} {et~al}\mbox{.}(2009){Ohkubo}, {Nomoto}, {Umeda}, {Yoshida},
  \& {Tsuruta}}]{Ohkubo+09}
{Ohkubo} T., {Nomoto} K., {Umeda} H., {Yoshida} N., {Tsuruta} S., 2009, \apj,
  706, 1184

\bibitem[{{Omukai} \& {Palla}(2001)}]{OmukaiPalla03}
{Omukai} K., {Palla} F., 2001, \apjl, 561, L55

\bibitem[{{Omukai}, {Schneider} \& {Haiman}(2008){Omukai}, {Schneider}, \&
  {Haiman}}]{Omukai+08}
{Omukai} K., {Schneider} R., {Haiman} Z., 2008, \apj, 686, 801

\bibitem[{{Park} \& {Ricotti}(2012)}]{PR12}
{Park} K., {Ricotti} M., 2012, \apj, 747, 9

\bibitem[{{Peebles}(1993)}]{Peebles93}
{Peebles} P.~J.~E., 1993, {Principles of Physical Cosmology}. Princeton
  University Press

\bibitem[{{Petri}, {Ferrara} \& {Salvaterra}(2012){Petri}, {Ferrara}, \&
  {Salvaterra}}]{Petri+12}
{Petri} A., {Ferrara} A., {Salvaterra} R., 2012, \mnras, 422, 1690

\bibitem[{{Regan} \& {Haehnelt}(2009)}]{RH09b}
{Regan} J.~A., {Haehnelt} M.~G., 2009, \mnras, 393, 858

\bibitem[{{Ricotti}(2009)}]{Ricotti09}
{Ricotti} M., 2009, \mnras, 392, L45

\bibitem[{{Ricotti} \& {Ostriker}(2004)}]{RicOst04}
{Ricotti} M., {Ostriker} J.~P., 2004, \mnras, 352, 547

\bibitem[{{Ricotti}, {Ostriker} \& {Gnedin}(2005){Ricotti}, {Ostriker}, \&
  {Gnedin}}]{Ricotti+05}
{Ricotti} M., {Ostriker} J.~P., {Gnedin} N.~Y., 2005, \mnras, 357, 207

\bibitem[{{Salpeter}(1955)}]{Salpeter55}
{Salpeter} E.~E., 1955, \apj, 121, 161

\bibitem[{{Schleicher}, {Spaans} \& {Glover}(2010){Schleicher}, {Spaans}, \&
  {Glover}}]{SSG10}
{Schleicher} D.~R.~G., {Spaans} M., {Glover} S.~C.~O., 2010, \apjl, 712, L69

\bibitem[{{Sesana}, {Volonteri} \& {Haardt}(2007){Sesana}, {Volonteri}, \&
  {Haardt}}]{Sesana+07}
{Sesana} A., {Volonteri} M., {Haardt} F., 2007, \mnras, 377, 1711

\bibitem[{{Sethi}, {Haiman} \& {Pandey}(2010){Sethi}, {Haiman}, \&
  {Pandey}}]{Sethi+10}
{Sethi} S., {Haiman} Z., {Pandey} K., 2010, \apj, 721, 615

\bibitem[{{Shakura} \& {Sunyaev}(1973)}]{SS73}
{Shakura} N.~I., {Sunyaev} R.~A., 1973, \aap, 24, 337

\bibitem[{{Shang}, {Bryan} \& {Haiman}(2010){Shang}, {Bryan}, \&
  {Haiman}}]{SBH10}
{Shang} C., {Bryan} G.~L., {Haiman} Z., 2010, \mnras, 402, 1249

\bibitem[{{Shankar} {et~al}\mbox{.}(2010){Shankar}, {Crocce},
  {Miralda-Escud{\'e}}, {Fosalba}, \& {Weinberg}}]{Shankar+10}
{Shankar} F., {Crocce} M., {Miralda-Escud{\'e}} J., {Fosalba} P., {Weinberg}
  D.~H., 2010, \apj, 718, 231

\bibitem[{{Shankar} {et~al}\mbox{.}(2004){Shankar}, {Salucci}, {Granato}, {De
  Zotti}, \& {Danese}}]{Shankar+04}
{Shankar} F., {Salucci} P., {Granato} G.~L., {De Zotti} G., {Danese} L., 2004,
  \mnras, 354, 1020

\bibitem[{{Shankar}, {Weinberg} \& {Miralda-Escud{\'e}}(2009){Shankar},
  {Weinberg}, \& {Miralda-Escud{\'e}}}]{Shankar+09a}
{Shankar} F., {Weinberg} D.~H., {Miralda-Escud{\'e}} J., 2009, \apj, 690, 20

\bibitem[{{Shapiro}, {Iliev} \& {Raga}(1999){Shapiro}, {Iliev}, \&
  {Raga}}]{Shapiro+99}
{Shapiro} P.~R., {Iliev} I.~T., {Raga} A.~C., 1999, \mnras, 307, 203

\bibitem[{{Shapiro}, {Iliev} \& {Raga}(2004){Shapiro}, {Iliev}, \&
  {Raga}}]{Shapiro+04}
---, 2004, \mnras, 348, 753

\bibitem[{{Shapiro}(2005)}]{Shapiro05}
{Shapiro} S.~L., 2005, \apj, 620, 59

\bibitem[{{Somerville} \& {Kolatt}(1999)}]{SomerKolatt99}
{Somerville} R.~S., {Kolatt} T.~S., 1999, \mnras, 305, 1

\bibitem[{{Spaans} \& {Silk}(2006)}]{SpaansSilk06}
{Spaans} M., {Silk} J., 2006, \apj, 652, 902

\bibitem[{{Stacy}, {Greif} \& {Bromm}(2010){Stacy}, {Greif}, \&
  {Bromm}}]{Stacy+10}
{Stacy} A., {Greif} T.~H., {Bromm} V., 2010, \mnras, 403, 45

\bibitem[{{Stacy}, {Greif} \& {Bromm}(2012){Stacy}, {Greif}, \&
  {Bromm}}]{Stacy+12}
---, 2012, \mnras, 422, 290

\bibitem[{{Tanaka} \& {Haiman}(2009)}]{TH09}
{Tanaka} T., {Haiman} Z., 2009, \apj, 696, 1798

\bibitem[{{Tanaka} \& {Menou}(2010)}]{TM10}
{Tanaka} T., {Menou} K., 2010, \apj, 714, 404

\bibitem[{{Taniguchi}(2004)}]{Taniguchi04}
{Taniguchi} Y., 2004, Progress of Theoretical Physics Supplement, 155, 202

\bibitem[{{Tegmark} {et~al}\mbox{.}(1997){Tegmark}, {Silk}, {Rees},
  {Blanchard}, {Abel}, \& {Palla}}]{Tegmark+97}
{Tegmark} M., {Silk} J., {Rees} M.~J., {Blanchard} A., {Abel} T., {Palla} F.,
  1997, \apj, 474, 1

\bibitem[{{Tseliakhovich}, {Barkana} \& {Hirata}(2011){Tseliakhovich},
  {Barkana}, \& {Hirata}}]{Tseliak+11}
{Tseliakhovich} D., {Barkana} R., {Hirata} C.~M., 2011, \mnras, 418, 906

\bibitem[{{Turk}, {Abel} \& {O'Shea}(2009){Turk}, {Abel}, \&
  {O'Shea}}]{Turk+09}
{Turk} M.~J., {Abel} T., {O'Shea} B., 2009, Science, 325, 601

\bibitem[{{Turk} {et~al}\mbox{.}(2012){Turk}, {Oishi}, {Abel}, \&
  {Bryan}}]{Turk+12}
{Turk} M.~J., {Oishi} J.~S., {Abel} T., {Bryan} G.~L., 2012, \apj, 745, 154

\bibitem[{{Turner}(1991)}]{Turner91}
{Turner} E.~L., 1991, \aj, 101, 5

\bibitem[{{Ueda} {et~al}\mbox{.}(2003){Ueda}, {Akiyama}, {Ohta}, \&
  {Miyaji}}]{Ueda+03}
{Ueda} Y., {Akiyama} M., {Ohta} K., {Miyaji} T., 2003, \apj, 598, 886

\bibitem[{{Venkatesan}, {Giroux} \& {Shull}(2001){Venkatesan}, {Giroux}, \&
  {Shull}}]{Venkat+01}
{Venkatesan} A., {Giroux} M.~L., {Shull} J.~M., 2001, \apj, 563, 1

\bibitem[{{Volonteri}, {Haardt} \& {Madau}(2003){Volonteri}, {Haardt}, \&
  {Madau}}]{VHM03}
{Volonteri} M., {Haardt} F., {Madau} P., 2003, \apj, 582, 559

\bibitem[{{Volonteri} \& {Natarajan}(2009)}]{VN09}
{Volonteri} M., {Natarajan} P., 2009, \mnras, 400, 1911

\bibitem[{{Volonteri} \& {Rees}(2005)}]{VolRees05}
{Volonteri} M., {Rees} M.~J., 2005, \apj, 633, 624

\bibitem[{{Volonteri} \& {Rees}(2006)}]{VolRees06}
---, 2006, \apj, 650, 669

\bibitem[{{Wechsler} {et~al}\mbox{.}(2002){Wechsler}, {Bullock}, {Primack},
  {Kravtsov}, \& {Dekel}}]{Wechsler+02}
{Wechsler} R.~H., {Bullock} J.~S., {Primack} J.~R., {Kravtsov} A.~V., {Dekel}
  A., 2002, \apj, 568, 52

\bibitem[{{Whalen}, {Abel} \& {Norman}(2004){Whalen}, {Abel}, \&
  {Norman}}]{Whalen+04}
{Whalen} D., {Abel} T., {Norman} M.~L., 2004, \apj, 610, 14

\bibitem[{{Wheeler} \& {Johnson}(2011)}]{WheelerJohnson11}
{Wheeler} J.~C., {Johnson} V., 2011, \apj, 738, 163

\bibitem[{{Willott} {et~al}\mbox{.}(2010{\natexlab{a}}){Willott}, {Delorme},
  {Reyl{\'e}}, {Albert}, {Bergeron}, {Crampton}, {Delfosse}, {Forveille},
  {Hutchings}, {McLure}, {Omont}, \& {Schade}}]{Willott+10a}
{Willott} C.~J. {et~al.}, 2010{\natexlab{a}}, \aj, 139, 906

\bibitem[{{Willott} {et~al}\mbox{.}(2010{\natexlab{b}}){Willott}, {Albert},
  {Arzoumanian}, {Bergeron}, {Crampton}, {Delorme}, {Hutchings}, {Omont},
  {Reyl{\'e}}, \& {Schade}}]{Willott+10b}
---, 2010{\natexlab{b}}, \aj, 140, 546

\bibitem[{{Wise} \& {Abel}(2008)}]{WA08}
{Wise} J.~H., {Abel} T., 2008, \apj, 685, 40

\bibitem[{{Wolcott-Green}, {Haiman} \& {Bryan}(2011){Wolcott-Green}, {Haiman},
  \& {Bryan}}]{JWG+11}
{Wolcott-Green} J., {Haiman} Z., {Bryan} G.~L., 2011, \mnras, 418, 838

\bibitem[{{Yoo} \& {Miralda-Escud{\'e}}(2004)}]{YooMirald04}
{Yoo} J., {Miralda-Escud{\'e}} J., 2004, \apjl, 614, L25

\bibitem[{{Yoshida}, {Omukai} \& {Hernquist}(2008){Yoshida}, {Omukai}, \&
  {Hernquist}}]{Yoshida+08}
{Yoshida} N., {Omukai} K., {Hernquist} L., 2008, Science, 321, 669

\bibitem[{{Zhang}, {Fakhouri} \& {Ma}(2008){Zhang}, {Fakhouri}, \&
  {Ma}}]{Zhang+08}
{Zhang} J., {Fakhouri} O., {Ma} C., 2008, \mnras, 389, 1521

\bibitem[{{Zhang}, {Woosley} \& {Heger}(2008){Zhang}, {Woosley}, \&
  {Heger}}]{ZhangW+08}
{Zhang} W., {Woosley} S.~E., {Heger} A., 2008, \apj, 679, 639

\bibitem[{{Zhao} {et~al}\mbox{.}(2003){Zhao}, {Jing}, {Mo}, \&
  {B{\"o}rner}}]{ZhaoD+03}
{Zhao} D.~H., {Jing} Y.~P., {Mo} H.~J., {B{\"o}rner} G., 2003, \apjl, 597, L9

\bibitem[{{Zhao} {et~al}\mbox{.}(2009){Zhao}, {Jing}, {Mo}, \&
  {B{\"o}rner}}]{ZhaoD+09}
---, 2009, \apj, 707, 354

\end{thebibliography}

\end{document}